\documentclass[iopams]{iopart}
\usepackage{geometry}
\usepackage{graphics}
\usepackage{graphicx}
\usepackage{amssymb}
\usepackage{subfig}
\usepackage{xcolor}

\begin{document}
\title[Asymptotic approximants for the trajectory of light]{Accurate closed-form trajectories of light around a Kerr black hole using asymptotic approximants}
\author{Ryne J. Beachley$^1$, Morgan Mistysyn$^2$, Joshua A. Faber$^{1,4}$, Steven J. Weinstein$^{3,4}$, Nathaniel S. Barlow$^{1,4}$}
\address{$^1$ School of Mathematical Sciences, Rochester Institute of Technology, Rochester, NY 14623} 
\address{$^2$ Department of Industrial and Systems Engineering, Rochester Institute of Technology, Rochester, NY 14623} 
\address{$^3$ Department of Chemical Engineering, Rochester Institute of Technology, Rochester, NY 14623} 
\address{$^4$ Center for Computational Relativity and Gravitation, Rochester Institute of Technology, Rochester, NY 14623} 
\ead{nsbsma@rit.edu}

\begin{abstract}
Highly accurate closed-form expressions that describe the full trajectory of photons propagating in the equatorial plane of a Kerr black hole are obtained using asymptotic approximants.  This work extends a prior study of the overall bending angle for photons (Barlow, et al. 2017, \textit{Class. Quantum Grav.}, \textbf{34}, 135017).  The expressions obtained provide accurate trajectory predictions for arbitrary spin and impact parameters, and provide significant time advantages compared with numerical evaluation of the elliptic integrals that describe photon trajectories. To construct approximants, asymptotic expansions for photon deflection are required in various limits.  To this end, complete expansions are derived for the azimuthal angle as a function of radial distance from the black hole in the far-distance and closest-approach (pericenter) limits, and new coefficients are reported for the bending angle in the weak-field limit (large impact parameter).
 \\

\end{abstract}

\submitto{\CQG}  \maketitle

\section{Introduction}\label{sec:intro}

\begin{figure}[h!]
\begin{center}
\includegraphics[width=6in]{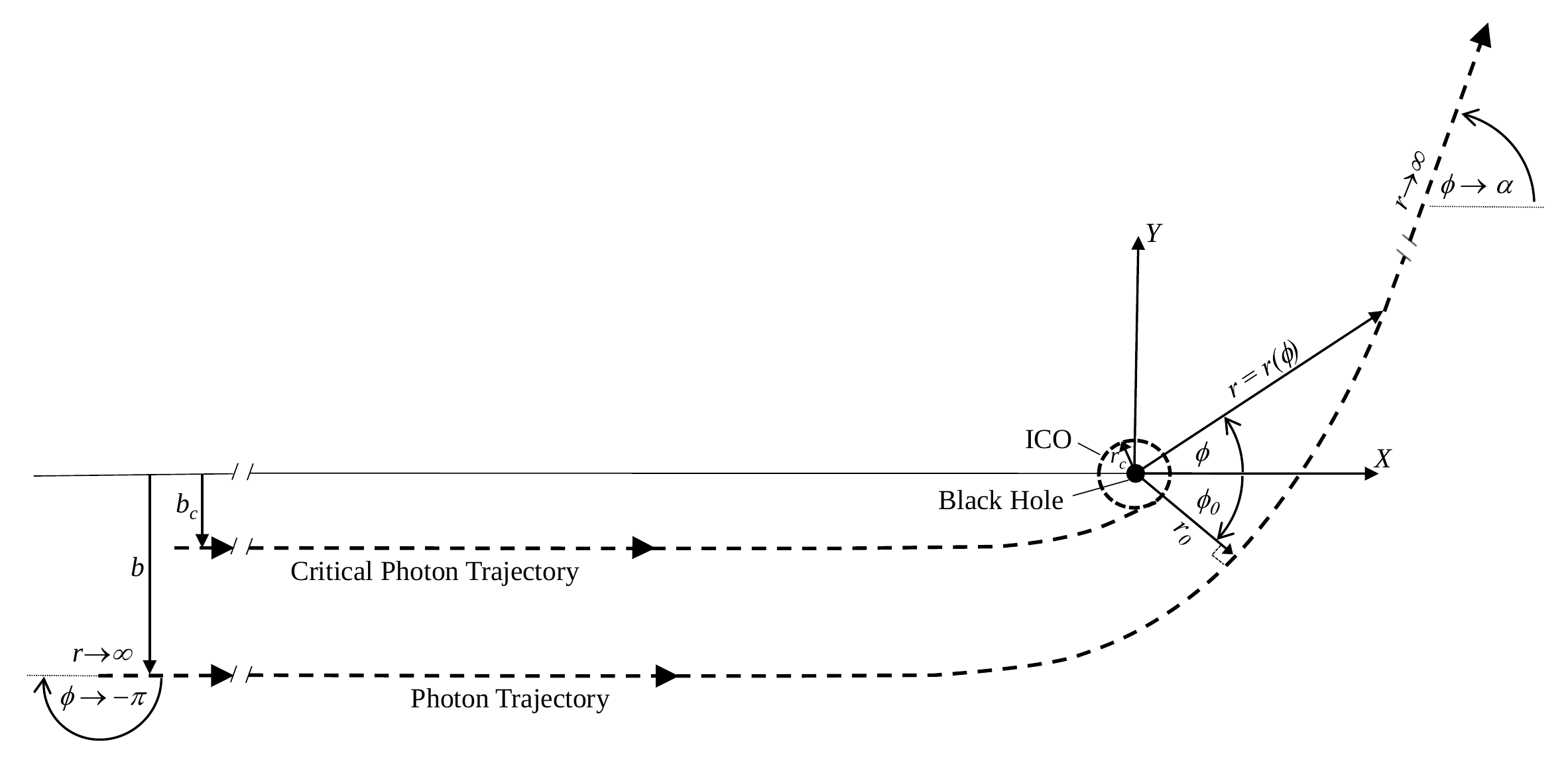}
 \end{center}
\caption{Schematic of photon trajectory in the equatorial plane of a Kerr black hole, parametrized as $r=r(\phi)$.  The $+Z$ direction is out of the page.  The labeled photon trajectory shows the relationship between the impact parameter, $b$, radial distance, $r$, azimuthal angle, $\phi$, and bending angle, $\alpha$.   The location of closest approach is denoted by coordinates $(r,\phi)=(r_0,\phi_0)$.  Also shown is the critical photon trajectory that corresponds to the Innermost Circular Orbit (ICO), showing the relationship between the critical impact parameter and radius, $b_c$ and $r_c$, respectively.}
\label{fig:definition}
\end{figure}

Light deflection in curved spacetimes is one of the earliest predictions of Einstein's general theory of relativity, and one of the best understood aspects of the theory.  The null geodesics describing photon trajectories have been investigated for a wide variety of physical configurations in a number of limits.  Mere years after general relativity was developed, the weak-field properties of light deflection around the sun were used to test the theory by Eddington and others during the eclipse of 1919. The limit where photons approach the innermost circular orbit (ICO\footnote{This applies for massless particles and occurs at different radii than the innermost stable circular orbit [ISCO] that pertains to massive bodies.}, see figure~\ref{fig:definition}), referred to as the strong-field limit, has also been explored for decades, tracing back to work by Hagihara  in the 1930's \cite{Hagihara:1930zz}, and further  studies by Darwin \cite{Darwin:1959zz} in the  following decades. After the initial construction of the Kerr metric describing spinning black holes~\cite{Kerr:1963ud}, many of the early results on null geodesics in these spacetimes were derived by Carter, beginning in the 1960's \cite{Carter:1968rr}.
We refer readers to Chandrasekhar's work on the subject for a thorough review on geodesics in black hole spacetimes \cite{Chandrasekhar:1983book}.

In our previous paper \cite{Barlow:2017b} (hereafter referred to as Paper 1), we examined the deflection of photons traveling in the equatorial plane of a Kerr (spinning) black hole.  Building off the work of Iyer and collaborators in particular \cite{Iyer:2006cn,Iyer:2009wa,Iyer:2009hq}, our work generated {\em approximate} expressions for the total light deflection, $\alpha$, as a function of the blackhole (henceforth abbreviated as BH) spin and impact parameter, $b$ (see figure~\ref{fig:definition}).  To do so, we utilized {\em asymptotic approximants}, i.e., closed-form expressions that connect asymptotic expansions obtained in different limits.  These approximants yielded highly accurate expressions for light bending.  Asymptotic approximants had previously been used to study thermodynamic phase behavior~\cite{Barlow:2012, Barlow:2014, Barlow:2015} and the solution of nonlinear boundary value problems~\cite{Barlow:2017}.

In Paper 1, we considered the {\em total} light deflection angle only, which describes the case where a photon begins at infinite (or quasi-infinite) separations from the black hole, passes by, and returns to infinite separations.  While gravitational lensing may be examined via the total bending angle, studies involving electromagnetic emission and absorption require knowledge of photon trajectories at finite distances from the BH.  As a particular example, one may consider an accretion disk around a black hole, a case which appears generically in astrophysics for BHs ranging in size from stellar-mass cases appearing as X-ray binaries up to Supermassive BHs observed as Active Galactic Nuclei.  The intense thermal heating resulting from the collisional process in these cases can yield copious high-energy emission, while the densities are potentially large enough in some regions that both emission and absorption must be considered to properly describe the full radiation transfer present.  In this work, we consider the \textit{full} trajectory of photons that traverse the equatorial plane of a Kerr BH starting and ending at spatial infinity.  We note that there is an immediate generalization for photons both emitted and absorbed at finite radius, $r$, as such photons superimpose onto trajectories that start and end at spatial infinity, and thus may be characterized by the parameters defined in figure~\ref{fig:definition}.

Calculations of light deflection in general have also been an active area of research, frequently within the context of understanding gravitational lensing and its potential observational signatures.  The complete solution for null geodesics in stationary BH spacetimes are expressed in terms of elliptic integrals (see \cite{Cadez:2004cg,Munoz:2014zz} for expressions involving Schwarzschild spacetimes, and \cite{Cadez:1998zz} for those involving Kerr).  Orbits in Kerr metrics and variants, including Kerr-de Sitter spacetimes with a cosmological constant, Kerr-Newman spacetimes with an electric charge, or both, have been computed exactly by Kraniotis  \cite{Kraniotis:2005zm,Kraniotis:2010gx,Kraniotis:2014paa} in terms of generalized multivariable hypergeometric functions and Weierstra{\ss} elliptic functions. The numerical evaluation of these integrals and/or complicated mathematical functions places a limit on the computational efficiency of ray-tracing codes that require large numbers of trajectories per volume.  Approximate schemes have been developed for both  Schwarzschild  \cite{Semerak:2014kra} and Kerr geodesics \cite{Aazami:2011tu,Aazami:2011tw}, but the tradeoffs between accuracy and computational efficiency are apparent.  Here, we report an approximate approach that provides a highly accurate closed-form expression that is also computationally efficient.

Our paper is organized as follows: In Section~\ref{sec:notation}, we review the basic equations governing null geodesics in Schwarzschild and Kerr spacetimes and establish notation, such as the critical impact parameter, $b_c$, and radius of closest approach, $r_0$, shown in figure~\ref{fig:definition}.  In Section~\ref{sec:limits}, we discuss the current understanding of light bending in terms of the strong-field ($b\to b_c$ in figure~\ref{fig:definition}) and weak-field ($b\to \infty$ in figure~\ref{fig:definition}) limits as a function of the BH spin.  Here we also review the necessary information from Paper 1 regarding the bending angle, provide new coefficients for the weak-field asymptotic expansion, introduce the closest-approach ($r\to r_0$ in figure~\ref{fig:definition}) and far-distance ($r\to\infty$ in figure~\ref{fig:definition}) limits in the radial direction, and provide a recursive formulation for the full asymptotic expansion in both of these limits. In Section~\ref{sec:approximant}, we use the method of asymptotic approximants to construct full analytic trajectories of photons that bridge the closest approach and far distance limits for given impact parameters, and in doing so, naturally bridge the strong and weak field limits.  In Section~\ref{sec:Results}, the asymptotic approximant is compared with the full numerical solution for the trajectory of light.  We conclude the main text of the paper in Section~\ref{sec:conclusions}, with a summary of key findings, possible future refinements, and a short discussion contrasting asymptotic approximants and asymptotic matching.  ~\ref{sec:Symmetry} provides an important symmetry relationship involving the photon trajectories that enables efficient calculations. ~\ref{sec:Recursion} provides useful series relations and the steps for developing recurrence relations for the series coefficients in both the asymptotic limits examined as well as the approximants.  For convenience,~\ref{sec:FifthOrderAlpha} provides a summary of coefficients needed to determine the overall bending angle ($\alpha$ in figure~\ref{fig:definition}) used in the construction of the approximant -- these coefficients are largely taken from Paper 1 with a few noted refinements.

\section{Light deflection: notation and conventions}\label{sec:notation}

We describe our photon trajectory in polar coordinates with radius $r$ and azimuthal angle $\phi$ (see figure~\ref{fig:definition}), defining the inverse radius $u\equiv r^{-1}$ for convenience.  Note that $r$ (and hence $u$) is taken here to be dimensionless, scaled by the distance unit $GM_{\rm BH}/c^2$ where $G$ is the gravitational constant, $M_{\rm BH}$ is the BH's mass, and $c$ is the speed of light. The BH spin parameter $a$  is dimensionless as well,  defined by 
\begin{eqnarray*}
a= cJ_{\rm BH}/[GM_{\rm BH}^2]
\end{eqnarray*}
where $J_{\rm BH}$ is the BH's angular momentum. The dimensionless spin parameter has a magnitude that naturally varies from $a=0$, a Schwarzschild BH without any angular momentum, up to $|a|=1$, an extremal Kerr BH.   Measuring BH spins observationally is difficult, but several groups have provided estimates.  For Sgr $A^*$, the supermassive black hole in the center of the Milky Way galaxy, estimates of the spin parameter range from $|a|=0.52$ from periodicities seen in infrared flares \cite{Genzel:2003as} up to spins of $a\approx 0.995$ 
as derived from periodicities in X-ray flares~\cite{Aschenbach:2004kj}.  Gravitational wave detectors like LIGO (the Laser Interferometer Gravitational-wave Observatory) and VIRGO will someday be able to constrain BH spins through their effects on the waveforms of merging binaries, but to date are only able to measure moderate misalignment of spins between the two BHs prior to merger, particularly in the source GW170104 \cite{Abbott:2017vtc}, without tight constraints on individual BH spins themselves.  In this work, we simply note that the range of physically motivated parameters extends from $|a|=0$ up to $|a|\sim 0.999$, with the upper limit still a matter of theoretical uncertainty.  The case $|a|=1$, which we also consider here, is included primarily for mathematical rather than astrophysically motivated reasons, as an asymptotic limit rather than a case likely to be observed in nature.

The Kerr geodesic equation for equatorial plane trajectories (thus having polar angle of $\pi/2$) is given by~\cite{Chandrasekhar:1983book,Iyer:2009wa}
\begin{eqnarray}
\nonumber
\frac{d\phi}{du} = \frac{1-2u(1-a/b)}{1-2u+a^2u^2}\frac{1}{\sqrt{h(u;a,b)}}\equiv f(u; a,b),~~\phi\in(-\pi,\phi_0)\\
h(u;a,b)=2\left(1-\frac{a}{b}\right)^2u^3-\left(1-\frac{a^2}{b^2}\right)u^2+\frac{1}{b^2},
\label{eq:int1}
\end{eqnarray}
where $\phi_0$ is the angle of closest approach and $\phi$ limits to $-\pi$ at $u=0$, as per figure~\ref{fig:definition}.  The non-dimensional impact parameter, $b$, is defined in terms of the $Z$-component of the photon's angular momentum, $L$, and its energy, $E$, as
\[b= c^2L/(EGM_{\rm BH}).\]
The relationship between $\phi$, $r$, and $b$ is shown in figure~\ref{fig:definition}.  As indicated, $b$ represents the minimum (perpendicular) distance between the BH center and the photon's unperturbed trajectory a photon would travel in the absence of gravitational curvature effects.

 We assume all photons travel with time, $t$, in a counterclockwise direction in the equatorial plane, with increasing azimuthal angle, $d\phi/dt >0$ (as indicated by the direction of the arrows along each trajectory in figure~\ref{fig:definition}).  For cases where the photon trajectory is prograde, we define the spin to be in the $+Z$ direction (out of the page in figure~\ref{fig:definition}), with $0<a\le 1$; for retrograde, the spin is in the $-Z$ direction and $-1\le a<0$. By integrating~\ref{eq:int1}, we find the azimuthal angle takes the form
\begin{eqnarray}
\phi&=&-\pi+\int_0^{u} \frac{1-2\hat{u}(1-a/b)}{[1-2\hat{u}+a^2\hat{u}^2]\sqrt{h(\hat{u};a,b)}}~d\hat{u}.
\label{eq:uintegral}
\end{eqnarray}
The denominator of the integrand in~(\ref{eq:uintegral}) has zeros that distinguish its solution; these are, in fact, integrable singularities except along the ICO.  Both the quadratic term and the cubic polynomial $h$ (defined in~(\ref{eq:int1})) have real-valued zeros, but there is one positive root of the cubic that occurs at a smaller value of $u$; it thus dominates the behavior of the integral.  The only exception to this situation is shown in Paper 1 for the extremal case $a=1$, where the zeros in the quadratic become coincident with those of $h$ as the ICO is approached, leading to a singular asymptotic behavior in that limit (see Paper 1: Appendix).  With these preliminary comments, we examine the nature of the cubic in more detail.

The cubic polynomial in $u$ that appears as $h(u;a,b)$ in~(\ref{eq:int1}) has three real roots, two of which are always positive and one negative.  We denote the smaller positive root as $u_0$; this is the root referred to above that dominates the integral behavior.  This quantity, defined as $u_0=1/r_0$, is the  the largest value of $u$ that a photon beginning from large distance ($u\rightarrow 0$) can achieve before reaching the closest approach along its trajectory (occurring at $r_0$ in figure~\ref{fig:definition}), at which point it will begin to recede from the black hole.  We note that if we convert $h(u;a,b)$ to $h(1/r;a,b)$ in~(\ref{eq:int1}) and multiply through by $r^3$, we arrive at a cubic in $r$.  The distance of closest approach, $r_0$, may be found in terms of $a$ and $b$ by solving the cubic equation $h(1/r;a,b)$=0:
\begin{eqnarray}
r_0(a,b) = [u_0(a,b)]^{-1} = \frac{2}{\sqrt{3}}\sqrt{b^2-a^2}\cos\left\{\frac{1}{3}
\cos^{-1}\left(-3\sqrt{3}\frac{(b- a)^2}
{(b^2-a^2)^{3/2}}\right) \right\}.
\label{eq:r0}
\end{eqnarray}
Eq.~(\ref{eq:r0}) implies that $r_0$ and $b$ are monotonically related.  Thus,  for a given spin, photon trajectories can be parameterized by either $r_0$ or $b$.

In Paper 1, we considered the relationship between a given impact parameter, $b$, and the total bending angle, $\alpha$, both situated at infinite radial distances from the BH as shown in figure~\ref{fig:definition}; it was possible to do so without the details of the trajectory, according to the formula
\begin{eqnarray}
\alpha = -\pi + 2\int_0^{u_0} f(u; a,b)~du.
\label{eq:int2}
\end{eqnarray}
Note that the integral in~(\ref{eq:int2}) evaluates to $\pi/2$ for a straight line trajectory in the {\em absence} of a gravitating source, yielding $\alpha=0$, consistent with the geometry in figure~\ref{fig:definition}.

As written,~(\ref{eq:int1}) and all further equations involving $\phi$ are valid in a portion of the domain in figure~\ref{fig:definition}, corresponding to $\phi\in(-\pi,\phi_0)$, for which $d\phi/dr<0$ (i.e., $d\phi/du>0$) and $r\in(\infty,r_0)$ (i.e., $u\in(0,u_0)$).  As shown in Paper 1, the remaining region corresponding to $\phi'\in(\phi_0,\alpha)$, and for which $d\phi'/dr>0$ (i.e., $d\phi'/du<0$), is described by~(\ref{eq:int1}) (replacing $\phi$ with $\phi'$) and with a right-hand side of $-f$.  This observation reveals that the full trajectory is symmetric about $r=r_0$ (or $u=u_0$).  To map out the trajectory of a photon, the procedure taken is to prescribe an impact parameter, $b$ (and thus, $r_0$), at $r=\infty$, which corresponds to $u=0$.  At any finite distance $r$ , the local angle $\phi$ may be determined from~(\ref{eq:uintegral}) where $u\le u_0$.  Once $\phi$ is obtained, the local ($X,Y$) coordinates of the trajectory are extracted as $X=r\cos(\phi)$ and $Y=r\sin(\phi)$ in the region $\phi\in(-\pi,\phi_0)$. The remaining portion of the trajectory in $\phi'\in(\phi_0,\alpha)$ can be mapped out using the symmetry condition of~(\ref{eq:A3}) in~\ref{sec:Symmetry}.  

\begin{figure}[h!]
\begin{center}
\subfloat{(a) \includegraphics[width=5.8in]{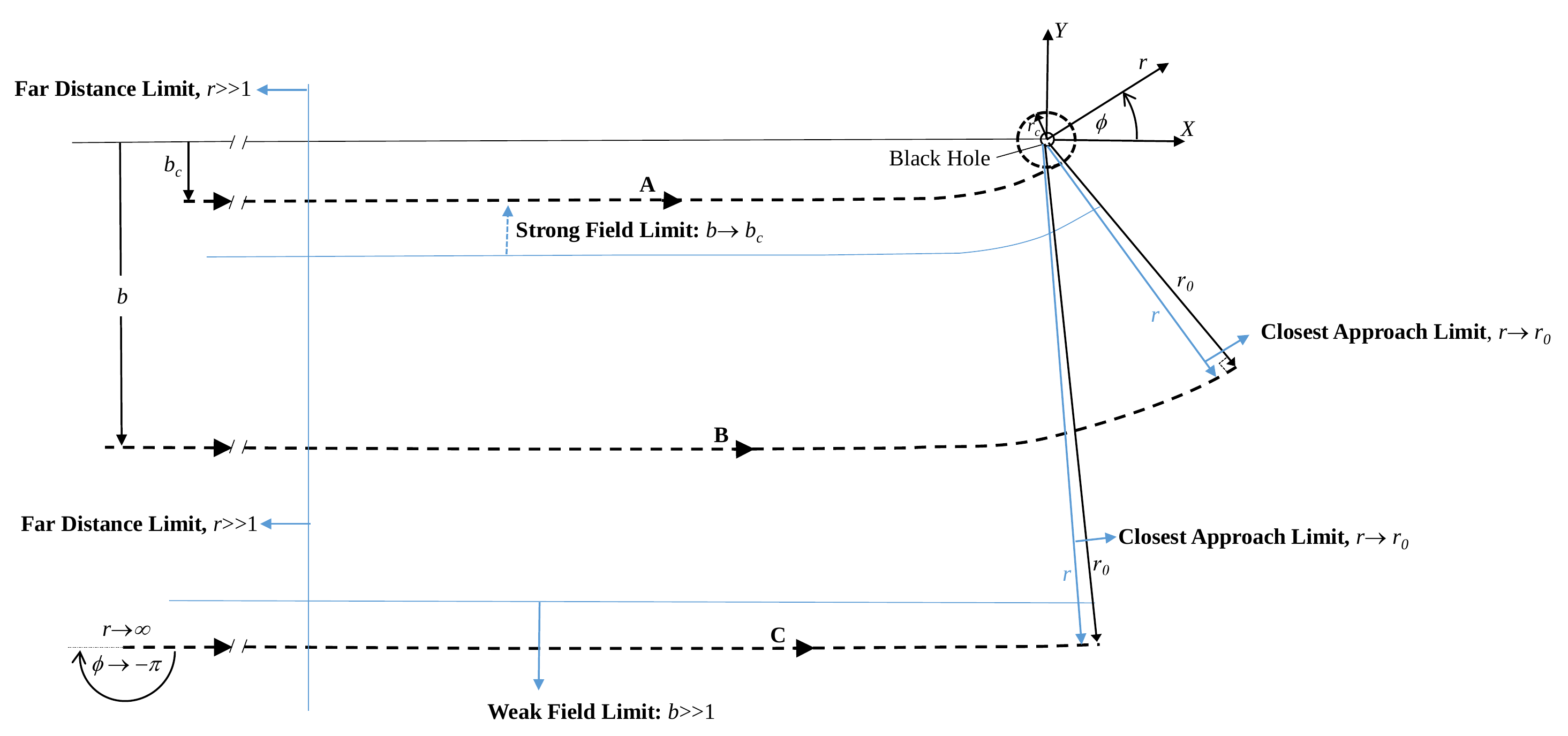}}\\
 \subfloat{(b) \includegraphics[width=3.4in, angle=0]{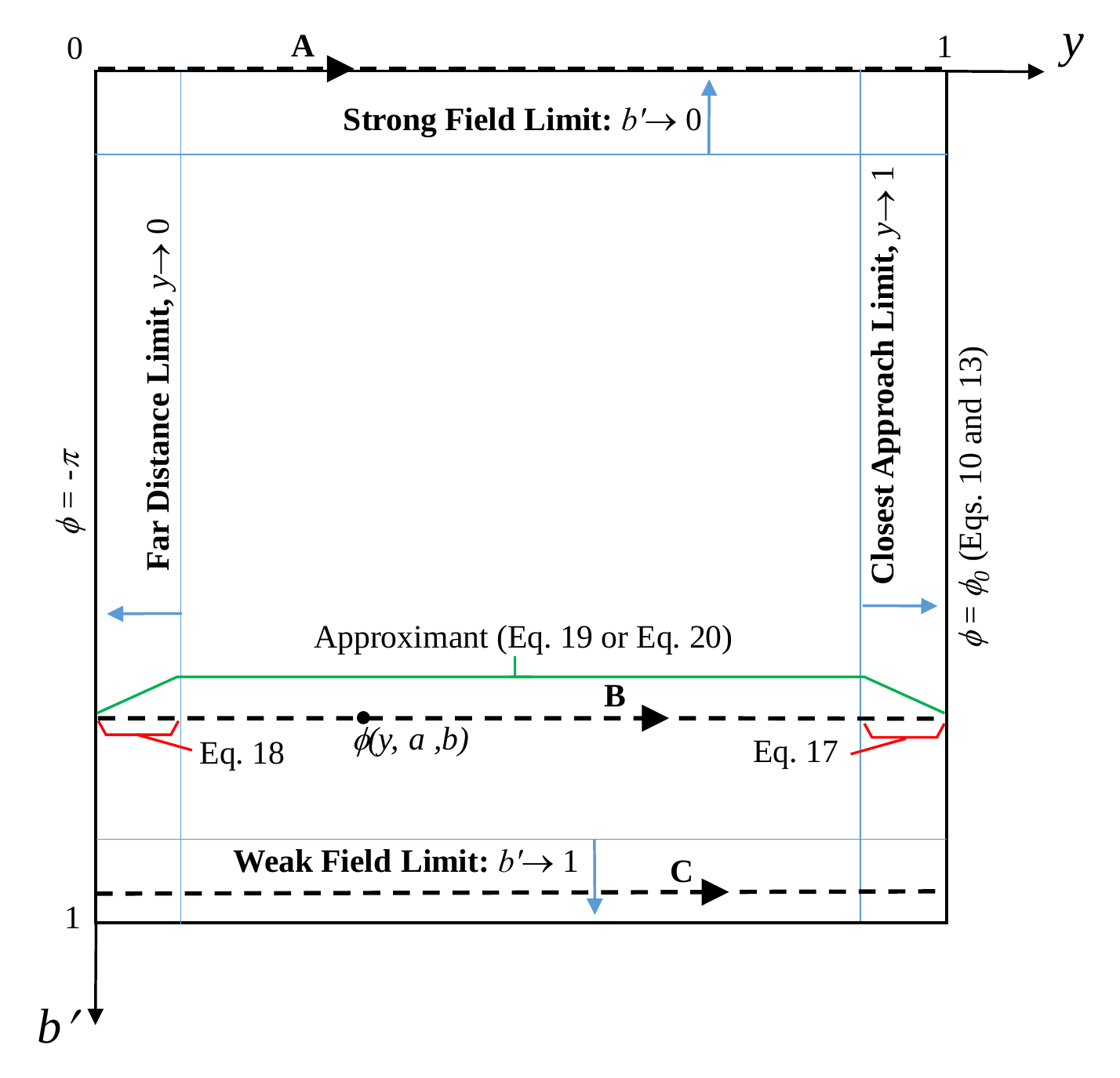}}
 \end{center}
\caption{Map of photon trajectories in the equatorial plane as they pass a Kerr BH, for any given spin. The space is characterized by 4 limits (described in Section~\ref{sec:limits}) indicated in (a) a portion of the trajectory of figure~\ref{fig:definition} in ($r,\phi$) coordinates for $-\pi\le\phi\le\phi_0$ and (b) in ($b',y$) coordinates where $b'=1-b_c/b$ and $y=r_0/r$. Trajectories in the physical plane (labeled A, B,C) in (a) correspond to those in the mapped domain in (b).}
\label{fig:schematic}
\end{figure}

In this work, we construct approximate solutions for the trajectory of photons as they pass a Kerr BH, as described by~(\ref{eq:int1}), such that the procedure described above can be carried out in an accurate but efficient manner.  Note that although a photon trajectory is characterized by the impact parameter that is defined at spatial infinity, photons themselves do not necessarily need to start or end at spatial infinity along a given trajectory.  The parameter space for such a trajectory in $(r,\theta)$ is encompassed within 4 limits, shown in figure~\ref{fig:schematic}a and each described in Section~\ref{sec:limits}.  Here we introduce the required notation and a convenient alternative coordinate system. In order to normalize the domain in $u$, we define a new quantity 
\begin{equation}
y\equiv u/u_0
\label{eq:y}
\end{equation}
to use as an integration variable, chosen so that the upper bound of the integral~(\ref{eq:uintegral}) satisfies $y=1$ in the limiting case and $0\le y\le 1$ in general.  We also adopt the same convention as \cite{Iyer:2006cn}, and normalize the impact parameter (shown in figure~\ref{fig:schematic}b) as
\begin{equation}
b' = 1-\frac{b_c}{b}
\label{eq:bp}
\end{equation}
 where $b_c$ (shown in figure~\ref{fig:schematic}a) corresponds to the limiting case of infinite bending angle, given by 
\begin{eqnarray}
b_c = 6\cos\left[\frac{1}{3}\cos^{-1}(-a)\right]-a
\label{eq:bc}
\end{eqnarray}
which follows from finding the $b$ value that minimizes $r_0$ in~(\ref{eq:r0})\footnote{This occurs at the value of $b$ at which the argument of the arccosine in~(\ref{eq:r0}) is $-1$, and is in general a solution of a cubic equation.}.  It also follows from~(\ref{eq:r0}) that the critical minimum separation, $r_c$ (shown in figure~\ref{fig:schematic}a), for a given spin, $a$, is 
\begin{eqnarray}
\nonumber
r_c(a) &=& 2+2\cos\left[\frac{2}{3}\cos^{-1}(-a)\right]\\
&=&3\frac{b_c-a}{b_c+a} = \sqrt{\frac{b_c^2-a^2}{3}},
\label{eq:rc}
\end{eqnarray}
where one may verify that $r_0 =r_c$ when $b=b_c$ in~(\ref{eq:r0}).  The effect of~(\ref{eq:y}) and~(\ref{eq:bp}) is to map the infinite physical domain of figure~\ref{fig:schematic}a to the unit square of figure~\ref{fig:schematic}b.  It is convenient to utilize the latter coordinate system and $b'$ notation when ultimately displaying results.  However, for purposes of notational clarity in our development, we will leave integrals and their expansions in terms of $b$. 

Applying the variable transformation~(\ref{eq:y}) to the integral~(\ref{eq:uintegral}), the azimuthal angle takes the following form in terms of $y$:
\begin{eqnarray}
\phi(y; a,b)&=&-\pi+\int_0^{y} \frac{u_0[1-2u_0 (1-a/b)\hat{y}]}{[1-2u_0\hat{y}+a^2u_0^2\hat{y}^2]\sqrt{h(u_0\hat{y};a,b)}}d\hat{y}\label{eq:intY}\nonumber\\
&\equiv&-\pi+\int_0^y g(\hat{y};a,b) ~d\hat{y}
\label{eq:intY}
\end{eqnarray}
which is linked to the bending angle, $\alpha$, considered in Paper 1 (and described by~(\ref{eq:int2})) via the expression
\begin{eqnarray}
 \phi(1; a,b)\equiv \phi_0(a,b)=\frac{\alpha-\pi}{2},
 \label{eq:bending}
\end{eqnarray}
where $\alpha=\alpha(a,b)$ and $\phi_0$ (the angle of closest approach) are both shown in figure~\ref{fig:definition}.  Our goal here is to determine approximate formulae for $\phi$ in terms of the three parameters $a$, $b$, and $y$.  As in Paper 1, we will treat the BH spin $a$ solely as a parameter. In contrast to that work, in which the impact parameter $b$ was generally treated as a variable, here we use $y$ as the primary expansion variable for our approximants.   There is physical motivation for this choice: a trajectory with fixed $b$ but variable $y$ represents the evolution of a photon through the BH space-time, as shown in the schematic in figure~\ref{fig:definition} (noting that $y=r_0/r$).

\section{Analytic limiting cases for light deflection}\label{sec:limits}

We now proceed to determine limiting asymptotic expressions for the azimuthal angle, $\phi$ (shown in figures~\ref{fig:definition} and~\ref{fig:schematic}), in various limits.  Note that expressions are derived in the mapped domain (figure~\ref{fig:schematic}b) in what follows, but the reader is may wish to refer back to the physical domain (figure~\ref{fig:schematic}a) to obtain a clear picture of the physical limits being examined. Also note that in what follows, the BH spin, $a$, is treated solely as a parameter, and is assumed fixed.

In Paper 1, we restricted attention to the overall deflection of the photon, $\alpha$, for a given scaled impact parameter, $b'$ (related to the usual impact parameter according to~(\ref{eq:bp}), as shown in figure~\ref{fig:definition}); this is related to the value of the azimuthal angle along the boundary $y=1$ (i.e., $\phi_0$ according to~(\ref{eq:bending})) as shown in figure~\ref{fig:schematic}b.  Our approximant for the bending angle $\alpha$ was formed by combining results from the \textit{strong-field limit}, corresponding to $b'\rightarrow 0$, and the \textit{weak-field limit}, corresponding to $b'\rightarrow 1$, and using asymptotic approximants to bridge the gap for intermediate values of $b'$.  As shown in figure~\ref{fig:schematic}b, there are two additional limits that correspond to the left and right-hand side of the domain, where now $b'$ is treated as a parameter.  In particular, the \textit{far-distance limit}, for which $y\rightarrow 0$, corresponds to cases where the BH trajectory begins at spatial infinity but ends arbitrarily far from the BH as well.  The opposite case is the \textit{closest-approach limit}, for which $y\rightarrow 1$ and the photon trajectory ends arbitrarily close to the pericenter passage.  Note that for large values of $b$ (i.e., $b'\to1$), the closest-approach may lie very far from the BH, while for trajectories with small values of $y$, the strong-field limit may apply for photons that are never actually close to the BH.  Regardless of the choice of the BH spin $a$, the azimuthal angle remains finite over the entire 2-dimensional phase space except for the strong-field, closest-approach case $b'=0$ and $y=1$ (upper-right corner of  figure~\ref{fig:schematic}b), which corresponds to the critical trajectory for which a photon asymptotically approaches the innermost circular orbit and never escapes the BH.

 While expansions can be developed for any of the 4 limits described above,  the behavior of the azimuthal angle in these various limits differs in important ways.  An effective canvassing of the space is needed to construct an asymptotic approximant, and  is accomplished using the particular expansions given in the following subsections.  Since the expansion variable, $y$, is the same as the integration variable in~(\ref{eq:intY}) for the far-distance and closest-approach limits, we choose to traverse the space of figure~\ref{fig:schematic}b by slicing horizontally.  This is also physically motivated, since each horizontal line in figure~\ref{fig:schematic}b corresponds to a photon trajectory, as shown in figure~\ref{fig:schematic}a.   The zeroth-order term of the closest-approach expansion is $\phi=\phi_0(a,b')$ (also shown in figure~\ref{fig:schematic}), which is not known exactly but an accurate expression (asymptotic approximant developed in Paper 1) is given in Section~\ref{sec:bending} below, which includes new terms (higher-order in $b'$) not disclosed previously.  All higher-order terms of the closest-approach expansion are determined exactly in Section~\ref{sec:closest}.  The zeroth-order term of the far-distance expansion is $\phi=-\pi$, as shown in figure~\ref{fig:schematic} (a or b).   All higher-order terms of the far-distance expansion are determined exactly in Section~\ref{sec:far}.  The approximant of Section~\ref{sec:approximant} is formed using the expansions given in Sections~\ref{sec:bending} through~\ref{sec:far}, and thus naturally incorporates all 4 limits of figure~\ref{fig:schematic}b.

\subsection{The closest approach limit: zeroth-order approximation ($y=1$, i.e., $r=r_0$). \label{sec:bending}}
Many elements of the zeroth-order closest-approach approximation have been examined in previous work~\cite{Barlow:2017b,Iyer:2009hq}, so here we summarize key results and extensions needed for the current work.  Along the closest-approach boundary ($y=1$, see figure~\ref{fig:schematic}b, rightmost vertical line), $\phi\equiv\phi_0=(\alpha-\pi)/2$ where the bending angle, $\alpha$, is completely described by the weak ($b'\to1$) and strong ($b'\to0$) field limits~\cite{Barlow:2017b,Iyer:2009hq}; these are obtained through expansions of~(\ref{eq:intY}) with $y=1$.  The weak field expansion (along $y$=1) is given to 4$^\mathrm{th}$ order in~\cite{Iyer:2009hq} and is extended in the current work to 7$^\mathrm{th}$ order as:
\begin{eqnarray}
\alpha&=&\sum_{n=1}^\infty a_n(b'-1)^n,\nonumber\\ \nonumber
a_1 &=& -\frac{4}{b_c}\\\nonumber
a_2 &=&\frac{-4a+15\pi/4}{b_c^2}\\\nonumber
a_3 &=& \frac{-4a^2+10\pi a-128/3}{b_c^3}\\\nonumber
a_4 &=& \frac{-4a^3+285\pi a^2/16-192a+3465\pi/64}{b_c^4}\\\nonumber
a_5 &=& \frac{-4a^4+27\pi a^3-512a^2+693\pi a/2-3584/5}{b_c^5}\\\nonumber
a_6 &=& \frac{-4a^5+1195\pi a^4/32-3200a^3/3+79695\pi a^2/64-17920a/3+255255\pi/256}{b_c^6}\\\nonumber
a_7 &=& \frac{-4a^6+195\pi a^5/4-1920a^4+13365\pi a^3/4-27136a^2+328185\pi a/32-98304/7}{b_c^7}\\
\vdots
\label{eq:alpha1}
\end{eqnarray}
while in the strong-field limit at $y$=1, the bending angle depends on the impact parameter according to the expression derived in Paper 1 as
\begin{equation}
 \alpha\sim-\pi+\beta+\gamma\ln\zeta+\delta_{a,1}\frac{\sqrt{3}}{b'}-\gamma\ln b'+O(b'\ln b'),~~\delta_{a,1} = \left\{
     \begin{array}{ll}
    0 &:~ a\neq1\\
       1&:~ a=1
     \end{array}
   \right..
 \label{eq:alpha0}
\end{equation}
where $\beta$, $\gamma$, and $\zeta$ are functions of $a$ given in~\cite{Barlow:2017b} (repeated in~\ref{sec:FifthOrderAlpha} for completeness).  An asymptotic approximant for $\alpha$ that bridges limits~(\ref{eq:alpha1}) and~(\ref{eq:alpha0})  was derived in Paper 1 as
\begin{equation}
 \alpha_{{\rm A},M}=-\pi+\beta+\gamma\ln\zeta+\delta_{a,1}\frac{\sqrt{3}}{b'}-\gamma\ln b'+\sum_{n=1}^{M+1}B_nb'^{\frac{n}{2}}\left(\Delta_{n+1}\sqrt{b'}\ln b'+\Delta_n\right)
 \label{eq:approximantAlpha}
\end{equation}
where $\Delta_n=1+(-1)^n$ and the $B_n$ coefficients are computed such that  the expansions of $\alpha_{{\rm A},M}$ and $\alpha$ about $b'=1$ (given by~(\ref{eq:alpha1})) are identical to $M^\mathrm{th}$-order; this requires an $(M+1)\times (M+1)$ matrix inversion; see~\cite{Barlow:2017b} for details.  For all figures that follow, a 5$^{th}$-order approximant  $\alpha_{{\rm A},5}$ is used, whose coefficients are given in~\ref{sec:FifthOrderAlpha}.  Note that the expression for $\alpha$, and by extension $\phi_0$, given by~(\ref{eq:approximantAlpha}) serves as the lowest-order term in the closest approach ($y\to1$) expansion of $\phi$. 

\subsection{The closest approach limit: higher order corrections ($y\to1$, i.e., $r\to r_0$) \label{sec:closest}}
In section~\ref{sec:bending}, an expression for the bending angle $\alpha$ is provided, which is needed for the zeroth order (in $y$) closest-approach term, $\phi(1;a,b)\equiv \phi_0 (a,b)=(\alpha-\pi)/2$.  Higher order terms in the full asymptotic series in this limit are obtained as follows. The integral expression~(\ref{eq:intY}) is rewritten as:
\begin{eqnarray}
\nonumber
\phi(y;a,b)=-\pi+\int_0^y g(\hat{y};a,b)~d\hat{y}& =& -\pi+\int_0^1 g(\hat{y};a,b)~d\hat{y} - \int_y^1 g(\hat{y};a,b)~d\hat{y} \\
&=&\phi_0 (a,b) -  \int_y^1 g(\hat{y};a,b)~d\hat{y}.
\label{eq:cal}
\end{eqnarray}
An asymptotic series as $y\to1$ cannot be obtained directly through a Taylor expansion of the integrand in~(\ref{eq:cal}), as it is singular at $y=1$.  To extract the series, it is worth considering the cubic in $\hat{y}$ ($h(u_0\hat{y};a,b)$ given by~(\ref{eq:int1})) that appears in the square root in the denominator of $g(\hat{y};a,b)$.   We may directly factor out the $(\hat{y}-1)$ term to find
 \begin{eqnarray}
h(u_0\hat{y};a,b)=
 (\hat{y}-1)\left(2\left(1-\frac{a}{b}\right)^2u_0^3 \hat{y}^2 -\frac{1}{b^2}\hat{y} -\frac{1}{b^2}\right).
 \label{eq:cubic_1root}
 \end{eqnarray}
 The cubic factorization~(\ref{eq:cubic_1root}) is inserted into $g(\hat{y};a,b)$ (given in~(\ref{eq:intY})) to isolate the singular behavior of the integral and, after rearrangement, the integral in~(\ref{eq:cal}) may be written as
\begin{eqnarray*}
\int_{y}^1 g(\hat{y};a,b)~d\hat{y} = \int_{y}^1 \frac{u_0[b-2u_0 (b-a)\hat{y}]}{[1-2u_0\hat{y}+a^2u_0^2\hat{y}^2]
\left(1+\hat{y}-2\left(b-a\right)^2u_0^3 \hat{y}^2\right)^{1/2}} ~\frac{d\hat{y}}{(1-\hat{y})^{1/2}}.
\end{eqnarray*}
Next, we define $z\equiv (1-\hat{y})^{1/2}$, and rewrite the integral as
\begin{eqnarray}
\label{eq:CDLtransformed}
\int_{y}^1 g(\hat{y};a,b)~d\hat{y}=
\int_0^{\sqrt{1-y}} \mathcal{G}(z;a,b)~dz,\\ \nonumber
\mathcal{G}(z;a,b)=\frac{2u_0[b-2u_0 (b-a)(1-z^2)]}{[1-2u_0(1-z^2)+a^2u_0^2(1-z^2)^2][1+(1-z^2)-2\left(b-a\right)^2u_0^3 (1-z^2)^2]^{1/2}}.
\end{eqnarray}
This new integrand is regular as $z$ approaches zero, and it is clear by inspection that its Taylor series involves only powers of $z^2$ and thus
\begin{eqnarray*}
\int_{y}^1 g(\hat{y};a,b)~d\hat{y}=&\int_0^{\sqrt{1-y}} \mathcal{G}(z;a,b)~dz=&\\
&\int_0^{\sqrt{1-y}} \sum_{n=0}^\infty \tilde{C}_n z^{2n} ~dz=&\sum_{n=0}^\infty \frac{\tilde{C}_n}{2n+1} (1-y)^{n+\frac{1}{2}}.\\
\end{eqnarray*}
Note that the resulting series contains only odd half-integer powers of $(1-y)$.  The closest-approach limit may be written compactly as
\begin{equation}
\phi=\phi_0+\sqrt{1-y}\sum_{n=0}^\infty C_n (y-1)^n,~C_n=\frac{(-1)^{n+1}}{2n+1}\tilde{C}_n
\label{eq:CDL}
\end{equation}
where 
\[\tilde{C}_n=\sum_{k=0}^{n} \left(\sum_{j=0}^{k} P_j S_{k-j} \right) Q_{n-k},\] 
\[P_0=2u_0b-4u_o^2(b-a),~P_1=4u_0^2(b-a),~P_{n\ge2}=0,\]
\[S_{n>0} =-\frac{1}{s_0}\sum_{j=1}^{n}s_jS_{n-j},~S_0=1/s_0,\]
\[s_0=1+u_0(-2+a^2u_0),~s_1=2u_0-2a^2u_0^2,~s_2=a^2u_0^2,~s_{n\ge3}=0,\]
\[Q_{n>0}=\frac{1}{nq_0}\sum_{j=1}^{n}(\frac{j}{2}-n)q_jQ_{n-j},~Q_0=1/\sqrt{q_0},\]
and
\[q_0=2[1-u_0^3(b^2-2ab+a^2)],~q_1=4u_0^3(a-b)^2-1,~q_2=-2u_0^3(b-a)^2,~q_{n\ge3}=0.\]
The steps for obtaining the recursion above are given in~\ref{sec:Recursion:Closest}. 

\begin{figure*}[h!]
\begin{center}
(a)\subfloat{\includegraphics[width=2.75in]{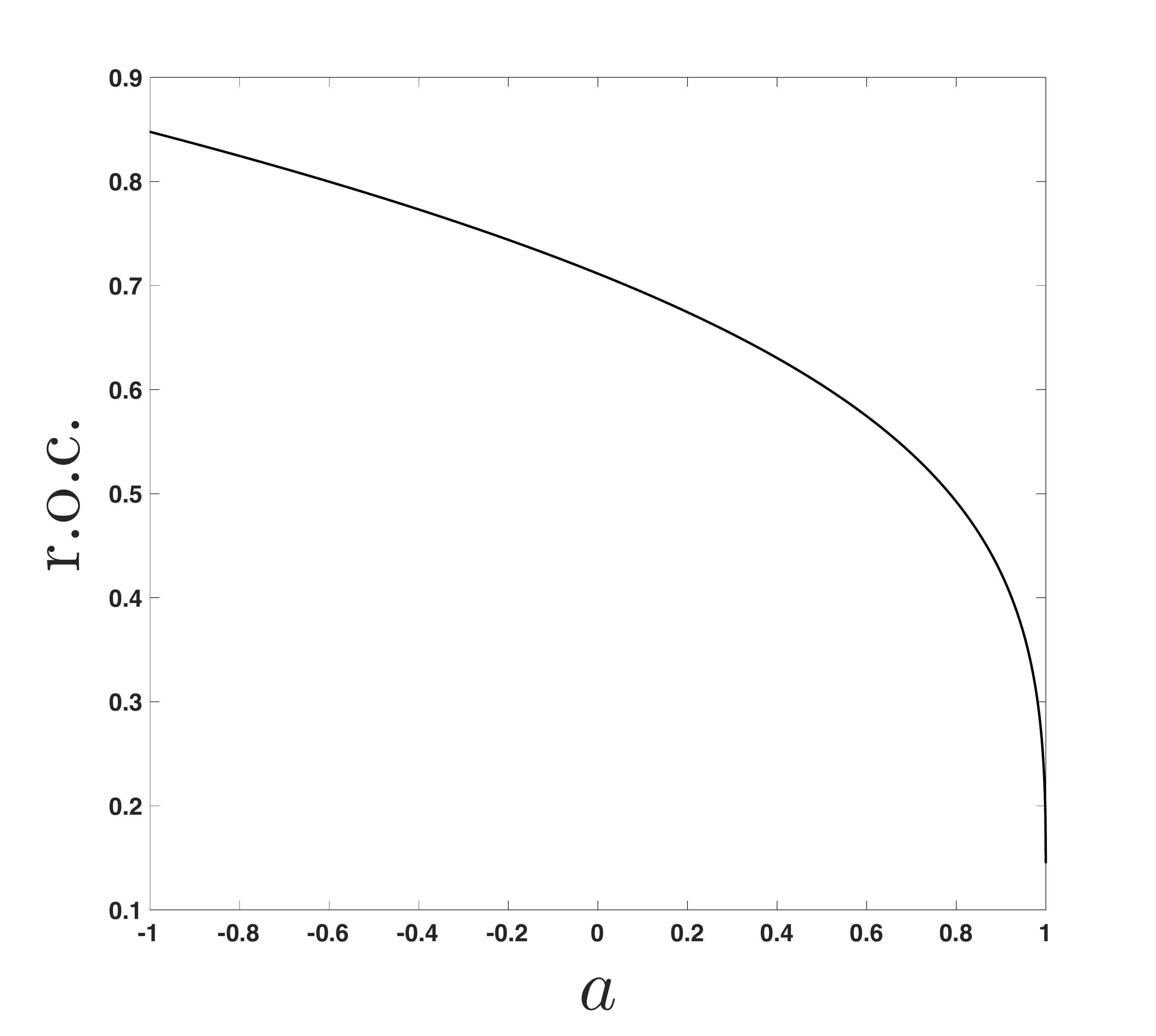}}
(b)\subfloat{\includegraphics[width=2.75in]{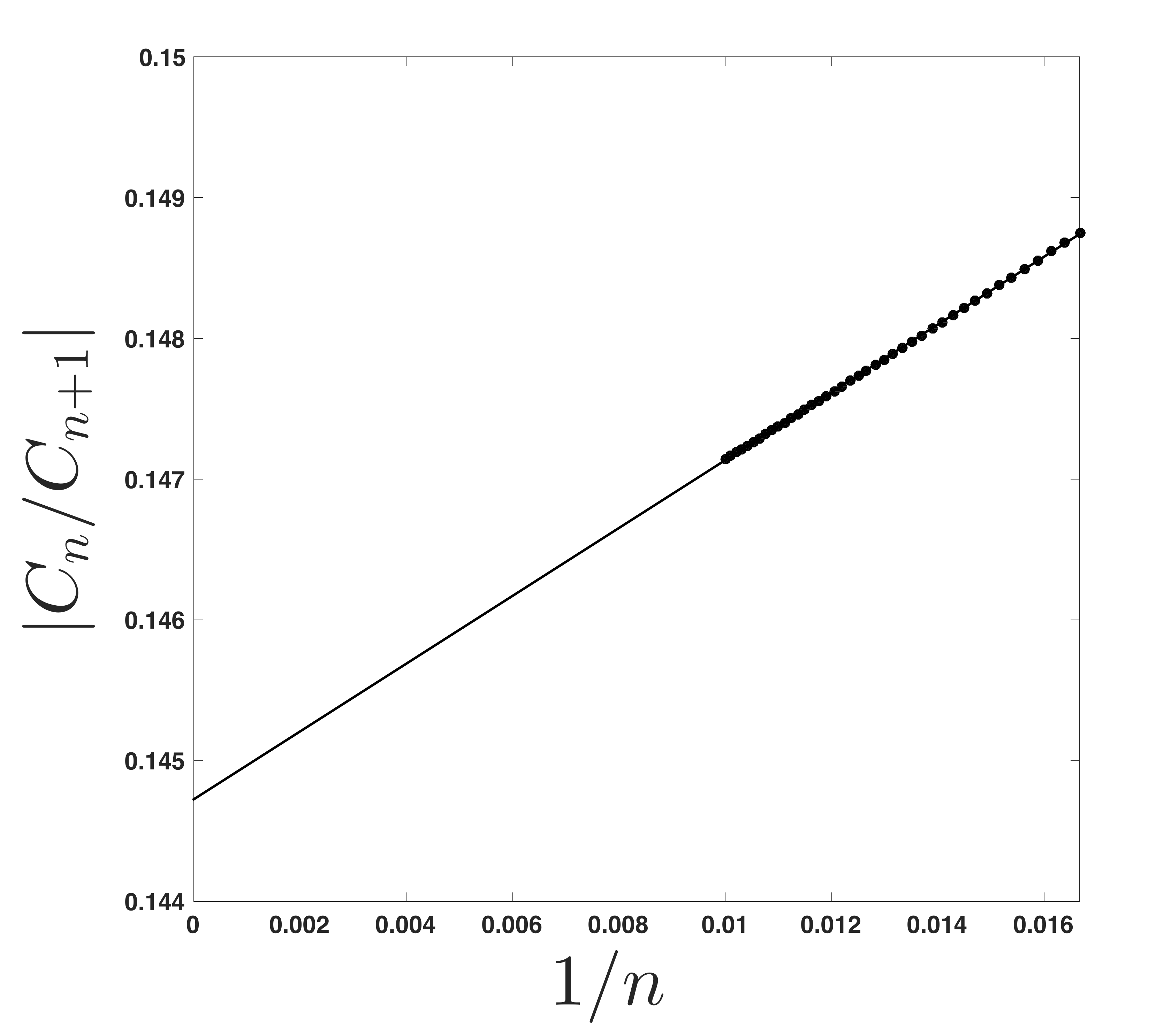}}
\end{center}
\caption{(a) Radii of convergence (r.o.c.) of the power series in~(\ref{eq:CDL}) as a function of $a$, for $b'\equiv1-b_c/b=0.1$. The r.o.c is extracted via examination of the singularities in~(\ref{eq:intY}) in the complex $\hat{y}$-plane.  (b)Domb-Sykes plot for the $C_n$ series in~(\ref{eq:CDL}) with $a=1$, $b'=0.1$, indicating a r.o.c. of $\approx$ 0.145.}
\label{fig:DS_CDL}
\end{figure*}

The radius of convergence (r.o.c.) of the power series in~(\ref{eq:CDL}) is prescribed by the distance from the closest singularity in the integrand of~(\ref{eq:intY}) from $\hat{y}=1$ in the complex $\hat{y}$-plane (excluding $\hat{y}=1$ itself, which has been factored out in~(\ref{eq:CDLtransformed})). Figure~\ref{fig:DS_CDL}a shows the r.o.c. as a function of $a$ for $b'$=0.1.  Since all terms of the closest-approach expansion are known recursively, the ratio-test may also be used to verify the r.o.c..  To this end, Domb-Sykes~\cite{DombSykes} plots of the reciprocal ratio of the ($n+1$) and $n^\mathrm{th}$ terms versus $1/n$ were constructed. Figure~\ref{fig:DS_CDL}b shows results for $a=1$ and $b'=0.1$, for 100 series terms.  It is found that the closest-approach series has a radius of convergence of $\approx$0.145, which agrees with the value in Figure~\ref{fig:DS_CDL}a at $a=1$.  Thus, the series in~(\ref{eq:CDL}) diverges at $y\approx1-0.145\approx0.855$ in this case.  It is noted that the r.o.c. of~(\ref{eq:CDL}) is indeed a function of $a$ and $b'$, but for all cases surveyed the Domb-Sykes results agree with the inspection of the singularities of~(\ref{eq:intY}), and the r.o.c. is indeed finite.

\subsection{The far-distance limit ($y\to0$, i.e., $r\to\infty$) \label{sec:far} }

The far distance limit consists of cases where the photon trajectory begins at spatial infinity and ends a distance for which the final value of $y$ can be considered infinitesimal (i.e., in figure~\ref{fig:schematic}a, $r$ remains large).
In this case our integral~(\ref{eq:intY}) takes a form that can be expanded in $y$:
\begin{eqnarray}
\phi(y;a,b)&=&-\pi+\int_0^y g(\hat{y};a,b)~d\hat{y}\nonumber\\
\nonumber\\
&=&-\pi+\int_0^y \left[\sum_{n=0}^\infty g_n\hat{y}^{n}\right] d\hat{y}=\sum_{n=0}^\infty \tilde{g}_ny^{n}, ~\tilde{g}_{n>0}=\frac{{g}_{n-1}}{n},~\tilde{g}_0=-\pi
\label{eq:fdl}
\end{eqnarray}
where
\[g_n=\sum_{j=0}^{n} \left(\sum_{k=0}^{j} p_k F_{j-k} \right) \tilde{c}_{n-j},\] 
\[p_0=u_0b,~p_1=-2u_0^2 (b-a),~p_{k>1}=0,\]
\[F_{n>0}=-\sum_{k=1}^nd_kF_{n-k},~F_0=1,\]
\[d_1=-2u_0,~d_2=a^2u_0^2,~d_{k>2}=0,\] 
\[\tilde{c}_{n>0}=\frac{1}{n}\sum_{k=1}^n\left(\frac{k}{2}-n\right)c_k\tilde{c}_{n-k},~\tilde{c}_0=1,\]
and
\[c_1=0,~c_2=-(b^2-a^2)u_0^2,~c_3=2(b-a)^2u_0^3,~c_{k>3}=0.\]
The steps for obtaining the recursion above are given in~\ref{sec:Recursion:Far}.  

\begin{figure*}[h!]
\begin{center}
(a)\subfloat{\includegraphics[width=2.75in]{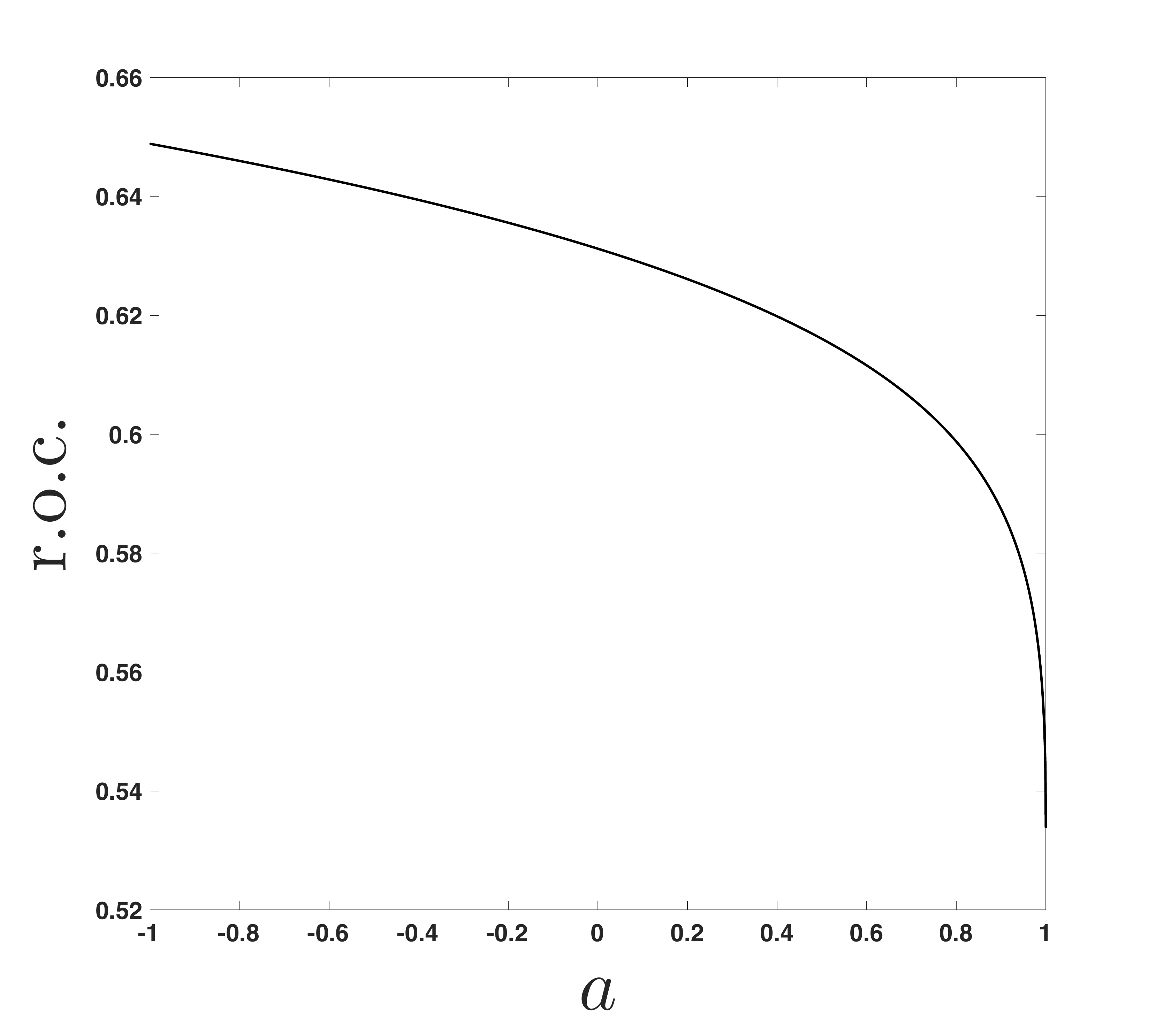}}
(b)\subfloat{\includegraphics[width=2.75in]{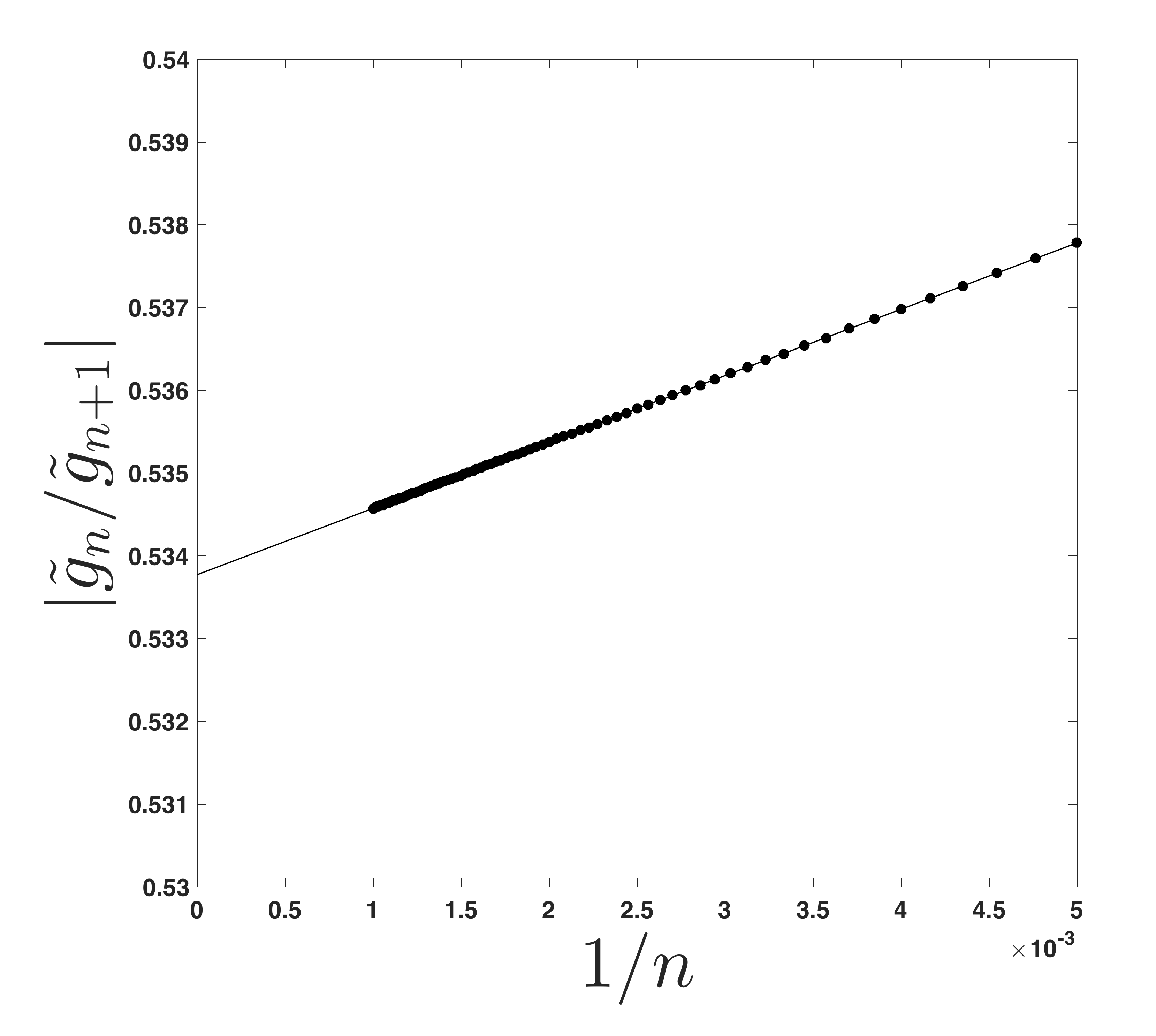}}
\end{center}
\caption{(a) Radii of convergence (r.o.c.) of the power series in~(\ref{eq:fdl}) as a function of $a$, for $b'\equiv1-b_c/b=0.1$. The r.o.c is extracted via examination of the singularities in~(\ref{eq:intY}) in the complex $\hat{y}$-plane. (b) Domb-Sykes plot for the $\tilde{g}_n$ series in~(\ref{eq:fdl}) with $a=1$, $b'=0.1$, indicating a r.o.c. of $\approx$ 0.534.}
\label{fig:DS_FDL}
\end{figure*}

The radius of convergence (r.o.c.) of the power series~(\ref{eq:fdl}) is prescribed by the distance from the closest singularity in the integrand of~(\ref{eq:intY}) from $\hat{y}=0$ in the complex $\hat{y}$-plane. Figure~\ref{fig:DS_FDL}a shows the r.o.c. as a function of $a$ for $b'$=0.1.  Since all terms of the closest-approach expansion are known recursively, the ratio-test may be used to verify the r.o.c..  To this end, Domb-Sykes~\cite{DombSykes} plots of the reciprocal ratio of the ($n+1$) and $n^\mathrm{th}$ terms versus $1/n$ are constructed. Figure~\ref{fig:DS_FDL}b shows results for $a=1$ and $b'=0.1$, for 1000 series terms.  The plot indicates that the far-distance series has a r.o.c. of $\approx$0.534, in agreement with the nearest singularity location to $y=0$ in figure~\ref{fig:DS_FDL}a, and thus diverges for $y$ values above this value. It is noted that the r.o.c. of~(\ref{eq:fdl}) is indeed a function of $a$ and $b'$, but for all cases surveyed the Domb-Sykes results agree with the inspection of the singularities of~(\ref{eq:intY}), and the r.o.c. is indeed finite.

\begin{figure*}[h!]
\begin{center}
\subfloat{\includegraphics[width=2.75in]{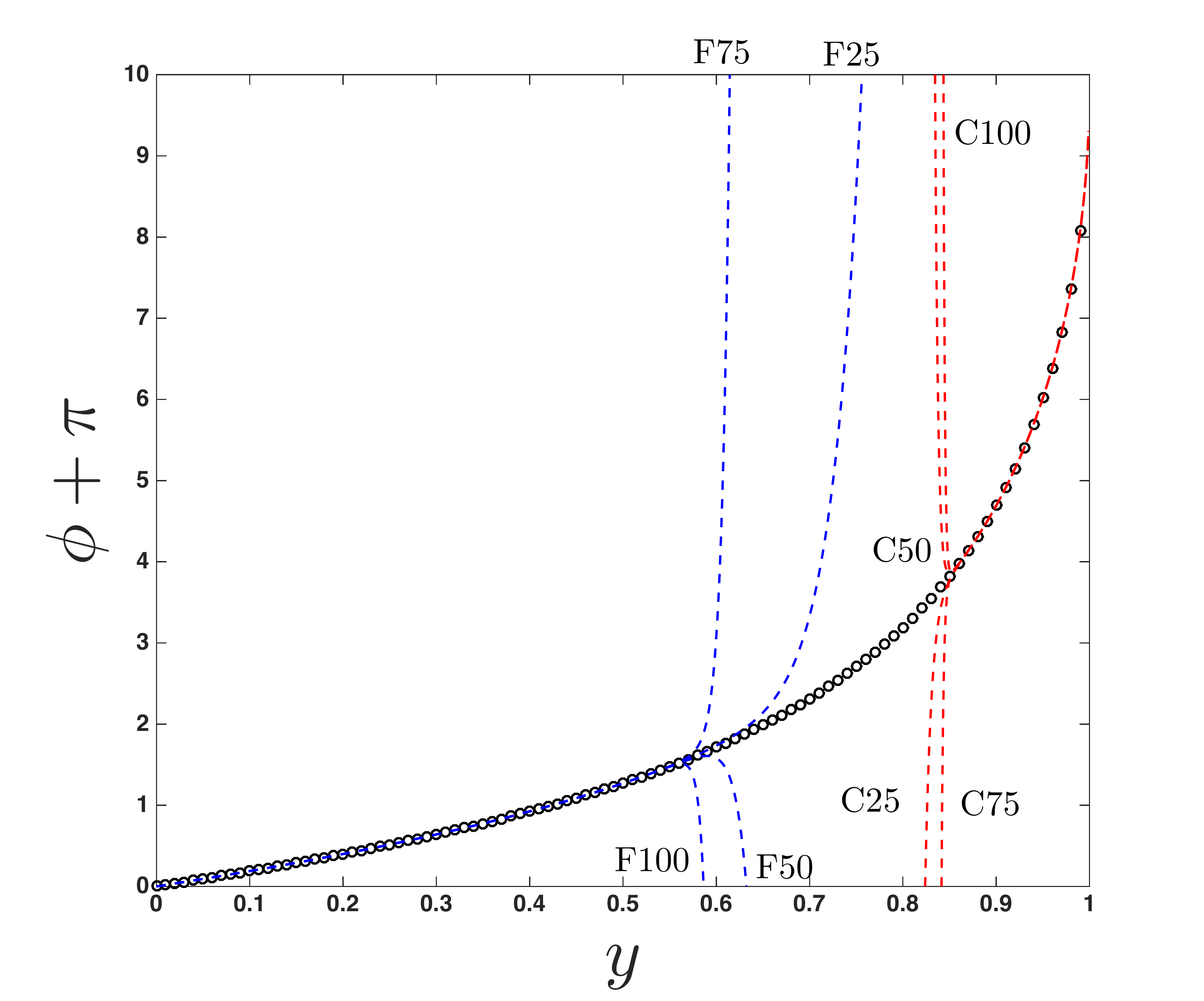}}
\end{center}
\caption{Photon angle $\phi$ vs. $y\equiv r_0/r$ for $a=1$ and $b'=0.1$.  The $N$-term far-distance expansion (denoted F$N$) given by~(\ref{eq:fdl}) and the closest-approach expansion using $N$-terms of the series in~(\ref{eq:CDL}) (denoted by C$N$) are compared with the numerical solution of~(\ref{eq:intY}) ($\circ$).  }
\label{fig:divergence}
\end{figure*}

The respective radii of convergence for the far-distance and closest-approach expansions restrict their direct use, as they are not convergent to determine $\phi$ over the domain $r_0<r<\infty$ (i.e., $0<y<1$).  For example, in the case of $a=1$ and $b'=0.1$, the closest-approach and far-distance series only converge in the respective intervals $0.855\lessapprox y\le1$ and $0\le y\lessapprox0.534$, based on the ratio tests illustrated in figures~\ref{fig:DS_CDL}b and~\ref{fig:DS_FDL}b.   These intervals of convergence are shown in figure~\ref{fig:divergence}, where each series is shown using 25, 50, 75, and 100 terms (dashed lines) and compared with the numerical solution of~(\ref{eq:intY}) ($\circ$).   Note that there is a gap for which neither series can be used to describe $\phi$.  Nevertheless, as shown in Section~\ref{sec:approximant}, these expansions may be used to develop a closed-form expression that is computationally fast, accurate, and describes photon trajectories in the full region $0<y<1$. 

%\newpage
\section{Asymptotic Approximant}\label{sec:approximant}

 To overcome the issue of divergent series such as those given in~(\ref{eq:fdl}) and~(\ref{eq:CDL}), there are a host of convergence acceleration (or ``re-summation'') techniques available that rely only on the original series itself (e.g. Pad\'e approximants, continued fractions,  Euler summation, etc.)~\cite{Bender}.  While such methods typically lead to an implementation improvement compared with the original series, global accuracy is not always guaranteed and the ``best'' re-summation technique is not always obvious~\cite{Clisby,Guerrero,Tan}.  In particular, any finite representation of an infinite power series can be analytically continued in a number of ways.  Additionally, summation techniques, such as Pad\'es (i.e. rational functions), can be effective because they approximate the pole singularities responsible for series divergence.   Of course, if the singularity responsible for the divergence is not a pole, the technique is not often as effective.  And, since the function we are trying to approximate is often unknown (only having its divergent series expansion to work with), it can thus be difficult to choose an efficient re-summation method a priori.  
 
 Asymptotic approximants constrain the analytic continuation of a series derived in one limit so that it can approach an asymptotic behavior in a different limit; an overview with examples is given in~\cite{Barlow:2017}. The accuracy of the approximant may be improved by including additional terms in the power series representation used in one of the limits.  The approach is validated by forming a convergent sequence of approximants as additional terms are added.  In the current work, the far-distance and closest-approach asymptotic expansions, given respectively by~(\ref{eq:fdl}) and~(\ref{eq:CDL}) are used to construct an approximant (see figure~\ref{fig:schematic}). 
   
The current problem mirrors a problem in thermodynamics that has been solved successfully using asymptotic approximants~\cite{Barlow:2014, Barlow:2015}, namely bridging a low-density power series at one end with a non-integer power law at the other end (the thermodynamic critical point).  The non-integer asymptotic behavior renders the rational-function form of standard Pad\'es innefective. In that problem, only knowledge of the \textit{leading-order} non-integer behavior is required to construct a uniformly accurate asymptotic approximant. Similarly, in the current problem, we have a regular power series in the far-distance limit and a non-integer asymptotic expansion in the closest-approach limit.  And akin to the thermodynamics problem,  we find that only the leading-order behavior of the closest-approach expansion is required to form a uniformly accurate approximant for any spin $a$ and most values of $b$ (higher-orders needed as $b\to b_c$), as is demonstrated in what follows (in Section~\ref{sec:Results}).   

An asymptotic approximant for the trajectory is constructed by satisfying the asymptotic expansions for the far-distance and closest-approach limits for a given impact parameter $b$ shown in figure~\ref{fig:schematic}a, corresponding to $b'$ in figure~\ref{fig:schematic}b.   The strong-field and weak-field limits are implicitly incorporated since their asymptotic forms are consistent with the closest-approach and far-distance limit expressions for small and large values of the impact parameter.  An approximant that satisfies the far-distance limit~(\ref{eq:fdl}) to $N^\mathrm{th}$-order and the $K$-term non-integer expansion in the closest-approach limit~(\ref{eq:CDL}) is given by
\begin{eqnarray}
\phi_{N,K}=\phi_0+\sqrt{1-y}\left\{\left[\sum_{n=0}^K C_n(y-1)^n\right]+(y-1)^{K+1}\sum_{n=0}^N A_n(y-1)^n\right\}, \nonumber\\
 A_n=\frac{1}{n!}\sum_{m=0}^N\left\{\frac{\Gamma(m+1)}{\Gamma(m-n+1)}\left[\sum_{j=0}^mT_{m-j}\frac{(-1)^{-K-1}~\Gamma(K+j+1)}{j!~\Gamma(K+1)}\right]\right\},\nonumber\\
 T_n=\frac{1}{\sqrt{\pi}}\left[\sum_{j=0}^n\tilde{\tilde{g}}_j\frac{\Gamma(n-j+1/2)}{\Gamma(n-j+1)}\right]-\frac{1}{n!}\sum_{j=0}^K\frac{(-1)^{j-n}~\Gamma(j+1)}{\Gamma(j-n+1)}C_j,\nonumber \\K=-1,0,1,2,\dots,
 \label{A}
 \end{eqnarray}
where $\phi_0$=$(\alpha-\pi)/2$, $\tilde{\tilde{g}}_0=\tilde{g}_0-\phi_0$, $\tilde{\tilde{g}}_{n>0}=\tilde{g}_{n>0}$ (defined in~(\ref{eq:fdl})), and $C_n$ coefficients are given in~(\ref{eq:CDL}).  For all approximant curves generated here, $\alpha$ (imbedded in $\phi_0=(\alpha-\pi)/2$) is computed using the bending angle approximant of Paper 1 taken to 5$^\mathrm{th}$ order, provided in Section~\ref{sec:bending} and with additional details in~\ref{sec:FifthOrderAlpha}.  The steps for obtaining the expressions for $A_n$ and $T_n$ in~(\ref{A}) are given in~\ref{sec:Recursion:Approximant}. The approximant~(\ref{A}) matches the closest-approach expansion (as $y\to1$) asymptotically to $(1/2+K)^{th}$-order (choosing $K\ge0$) and has an expansion about $y=0$ that is exactly the far distance series to $N^\mathrm{th}$-order.  Note that one may also set $K=-1$ to remove the $K$ series from~(\ref{A}).  This form of the approximant enforces only the zeroth-order closest approach limit $\phi_0$, but still matches the functional form of higher-order terms.  For $K=-1$, the approximant~(\ref{A}) reduces to 
\begin{eqnarray}
\phi_{N,-1}=\phi_0+\sqrt{1-y}\sum_{n=0}^N A_n(y-1)^n \nonumber \\
A_n=\frac{1}{n!\sqrt{\pi}}\sum_{m=0}^N\left\{\frac{\Gamma(m+1)}{\Gamma(m-n+1)}\left[\sum_{j=0}^m\tilde{\tilde{g}}_j\frac{\Gamma(m-j+1/2)}{\Gamma(m-j+1)}\right]\right\}.
 \label{approximant}
 \end{eqnarray}
 The reduced approximant given by~(\ref{approximant}) may be used to compute accurate photon trajectories within most of physical parameter space (i.e., that of figure~\ref{fig:schematic}b).  This is demonstrated in the cases presented in Section~\ref{sec:Results}, and guidance is then provided for when one might use the higher-order approximant given by~(\ref{A}) for $K\ge0$.  
 
 \section{Results and Discussion \label{sec:Results}}
\begin{figure*}[h!]
\begin{center}
\subfloat{(a) \includegraphics[width=2.75in]{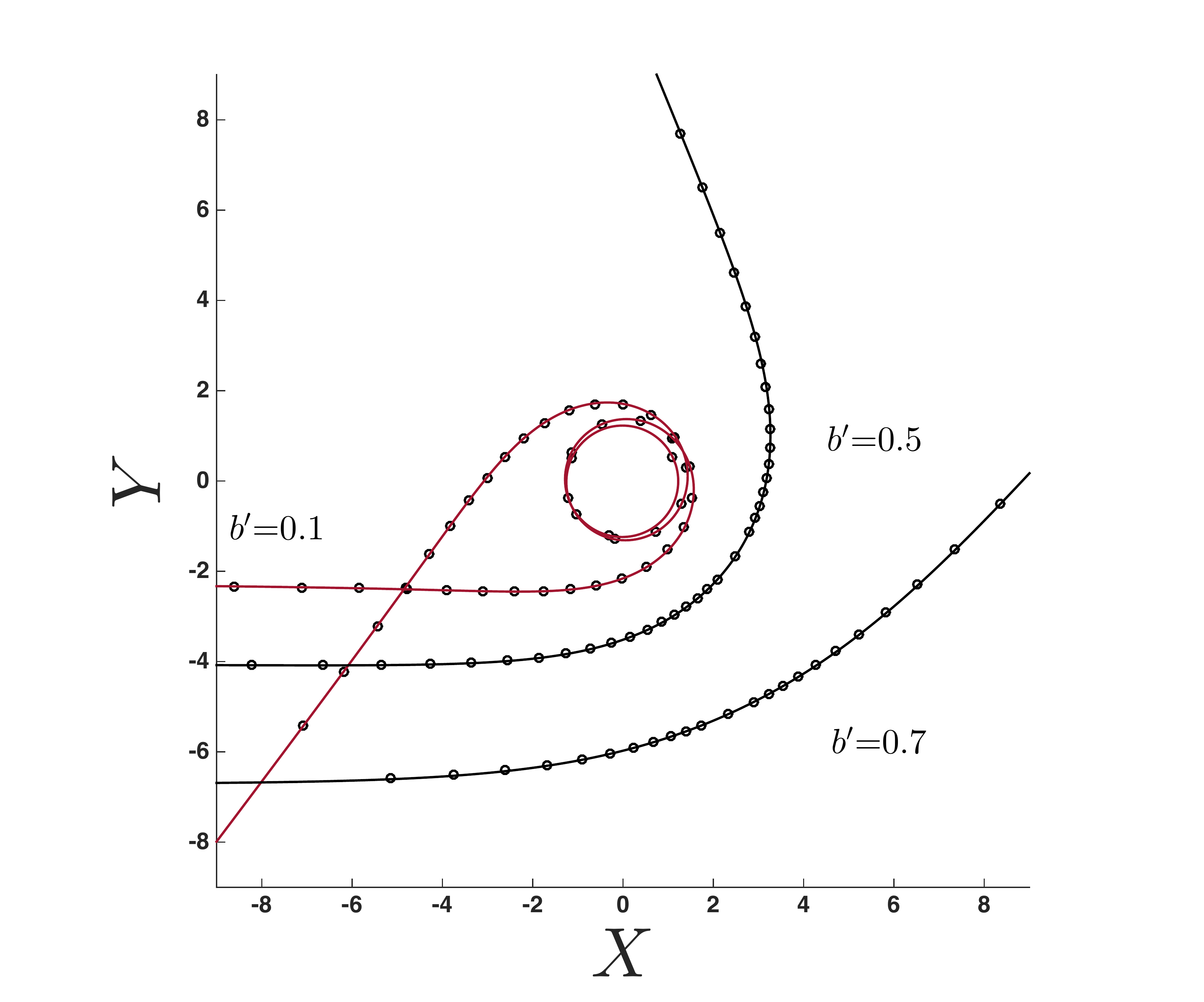}}
\subfloat{(b) \includegraphics[width=2.75in]{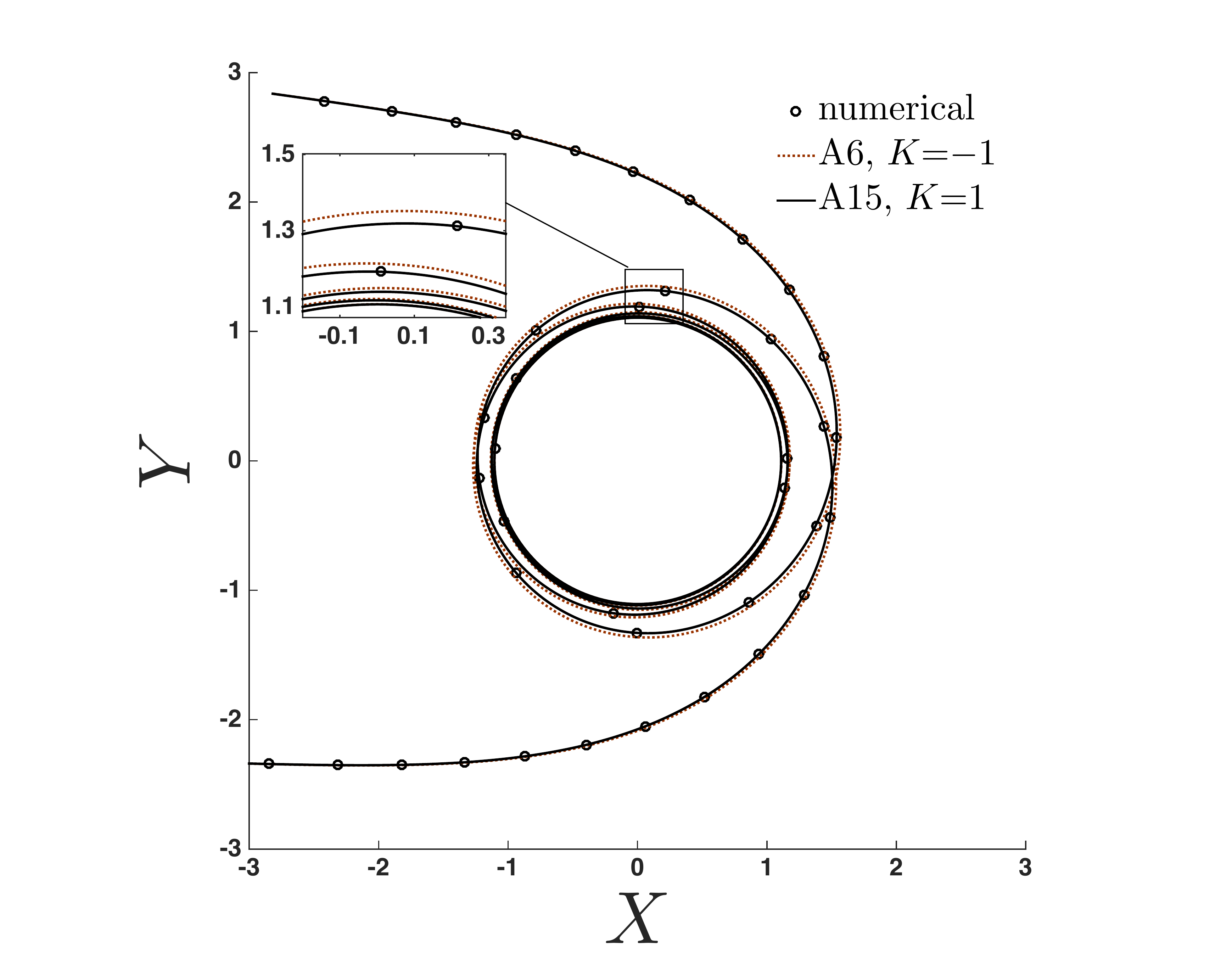}}\\
\subfloat{(c)\includegraphics[width=2.75in]{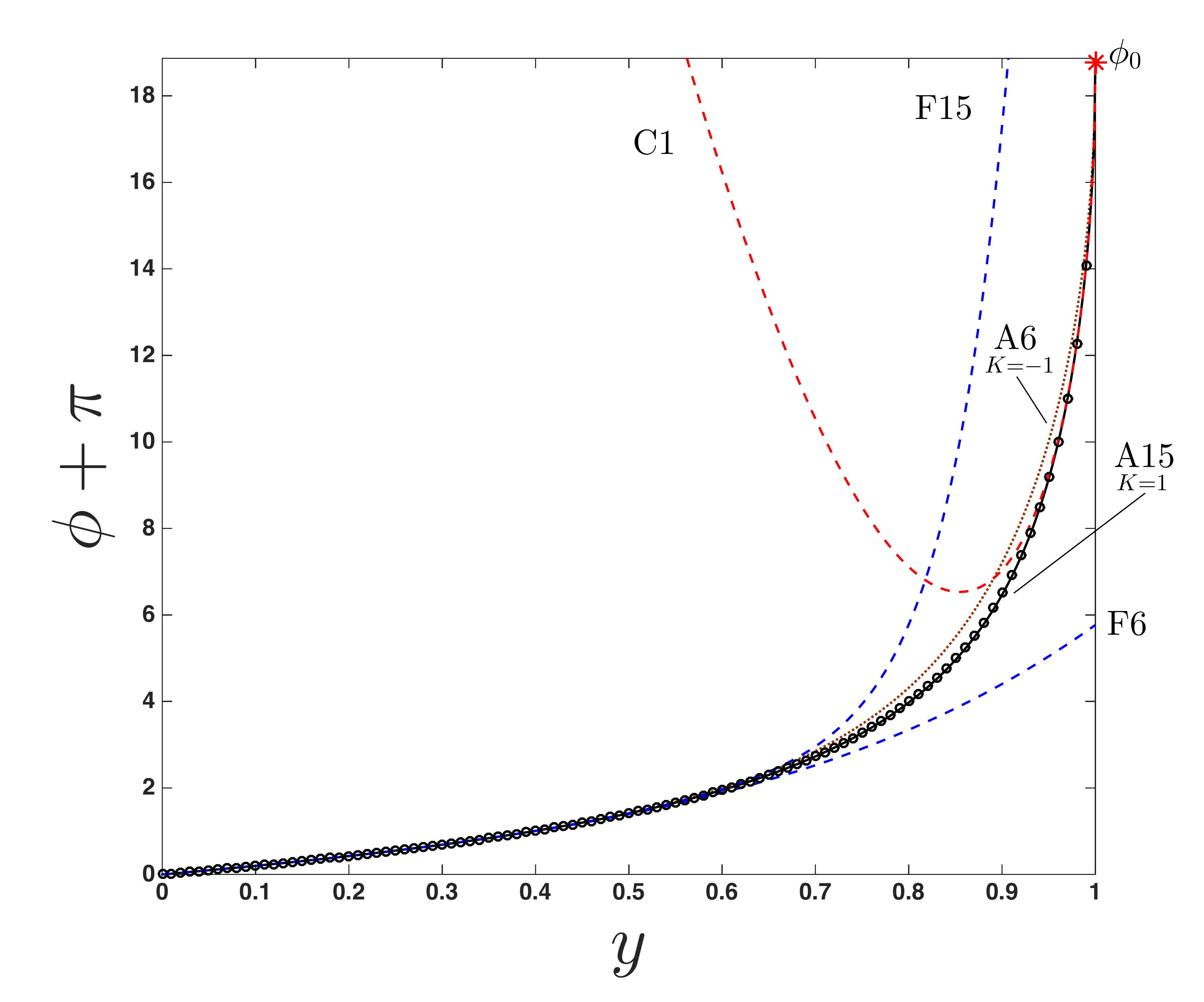}}
\subfloat{(d)\includegraphics[width=2.75in]{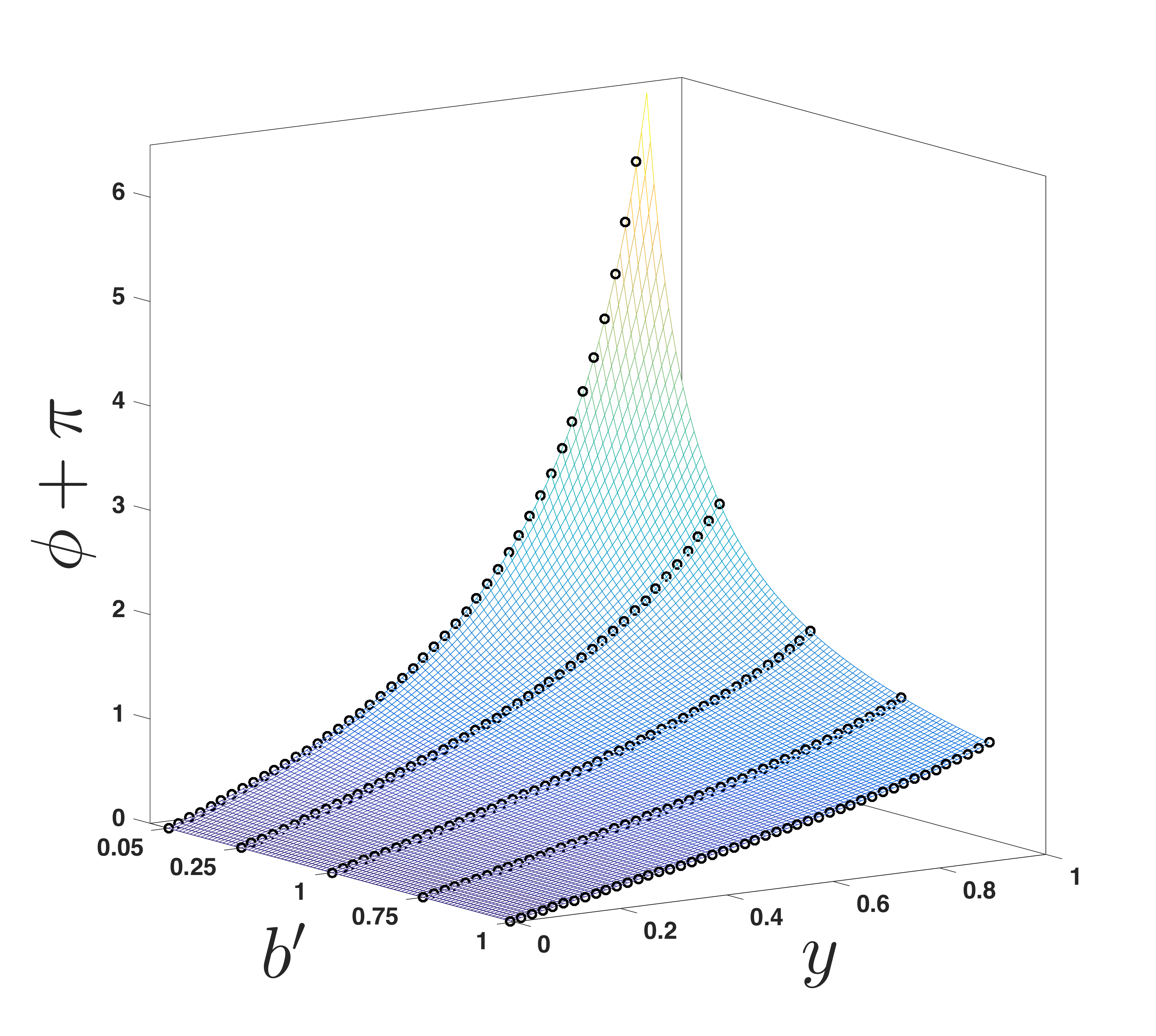}}
\end{center}
\caption{Trajectories of light around an extremal ($a$=1) Kerr BH, shown for different values of impact parameter $b'$, compared with the numerical solution ($\circ$) of~(\ref{eq:intY}). (a) The $K=-1$ approximant~(\ref{approximant}) (solid curves) using $N$=6. (b) Comparison between the $K=-1$ approximant~(\ref{approximant}) and $K=1$ approximant~(\ref{A}) for $b'=0.05$ and for different $N$, labeled A$N$.   (c) The approximant (using same naming conventions as (b)) shown in the mapped domain for $b'=0.05$, the $N$-term far-distance series~(\ref{eq:fdl}) (dashed curve, labeled F$N$), and the 3/2-order ($K=1$) closest-approach limit~(\ref{eq:CDL}) (dashed curve, labeled C1). (d) parameter space, as given by approximant~(\ref{A}) with $K$=1 and $N$=15.}
\label{fig:trajectory}
\end{figure*}

\begin{figure*}[h!]
\begin{center}
\subfloat{(a) \includegraphics[width=2.9in]{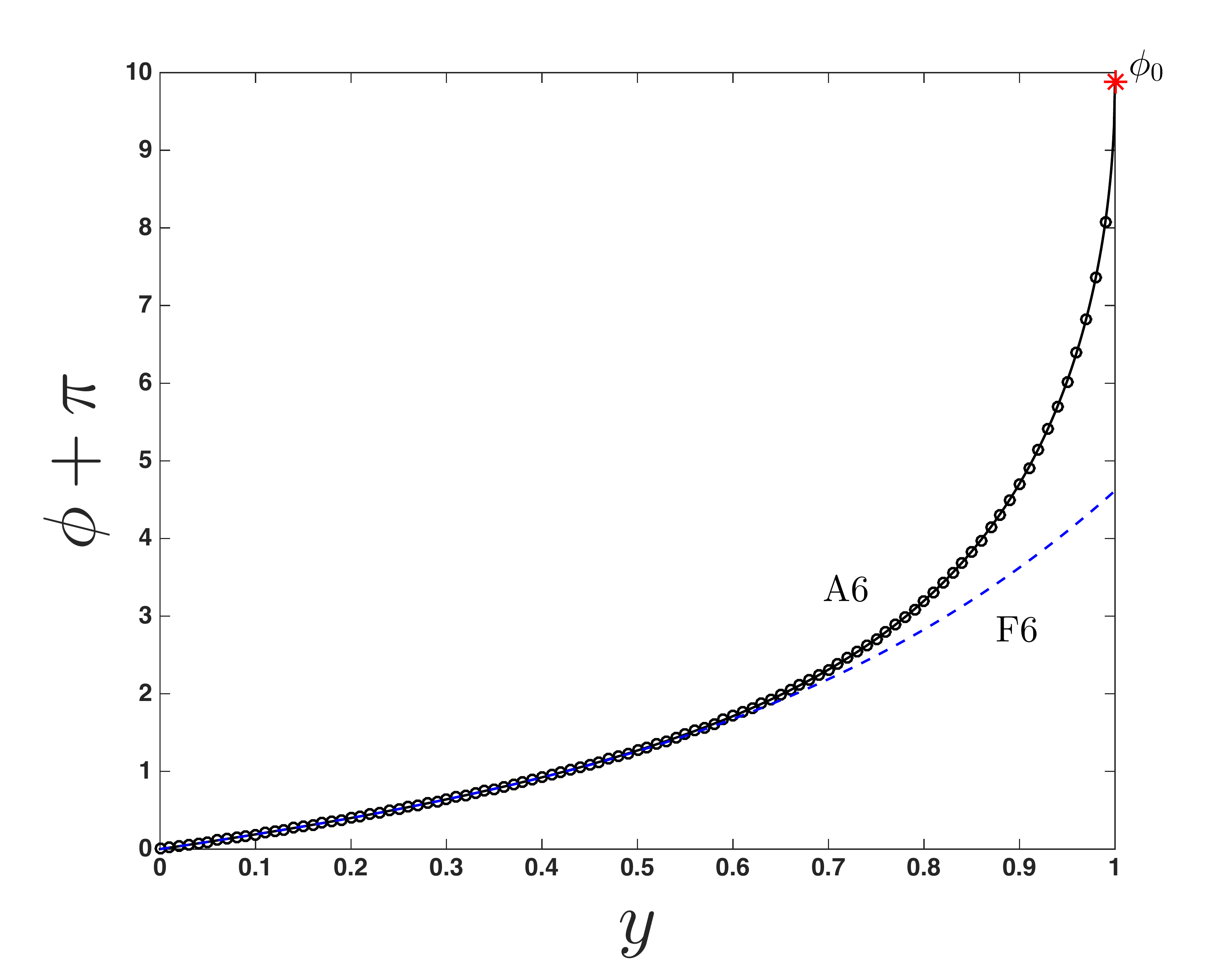}}
\subfloat{(b)\includegraphics[width=2.9in]{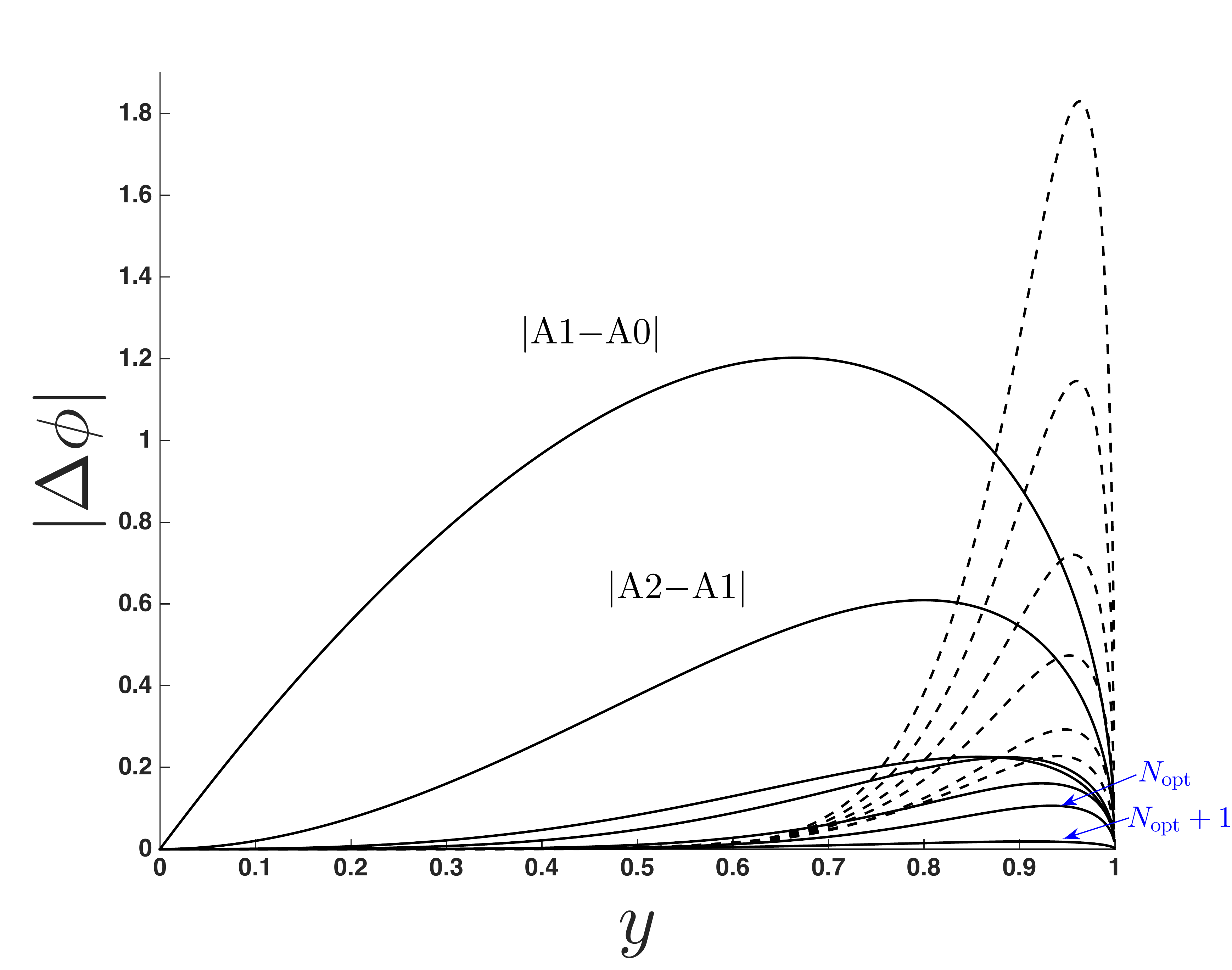}}
\end{center}
\caption{(a) Azimuthal angle, $\phi$, as a function of $y\equiv r_0/r$, for an extremal ($a$=1) Kerr BH with $b'$=0.1.  The approximant~(\ref{A}) (solid curve) (using $N$=6, $K=-1$) and the 6-term far-distance series used as an input to the approximant (dashed curve, labeled F6) are compared with the numerical solution ($\circ$) of~(\ref{eq:intY}).  (b) Convergence (solid curves moving downward) of the $K=-1$ approximant for increasing $N$, showing an optimal truncation of $N$=6 before divergent behavior sets in (dashed curves moving upward). }
\label{fig:extremalCauchy}
\end{figure*}

\begin{figure*}[h!]
\begin{center}
\subfloat{(a) \includegraphics[width=2.8in]{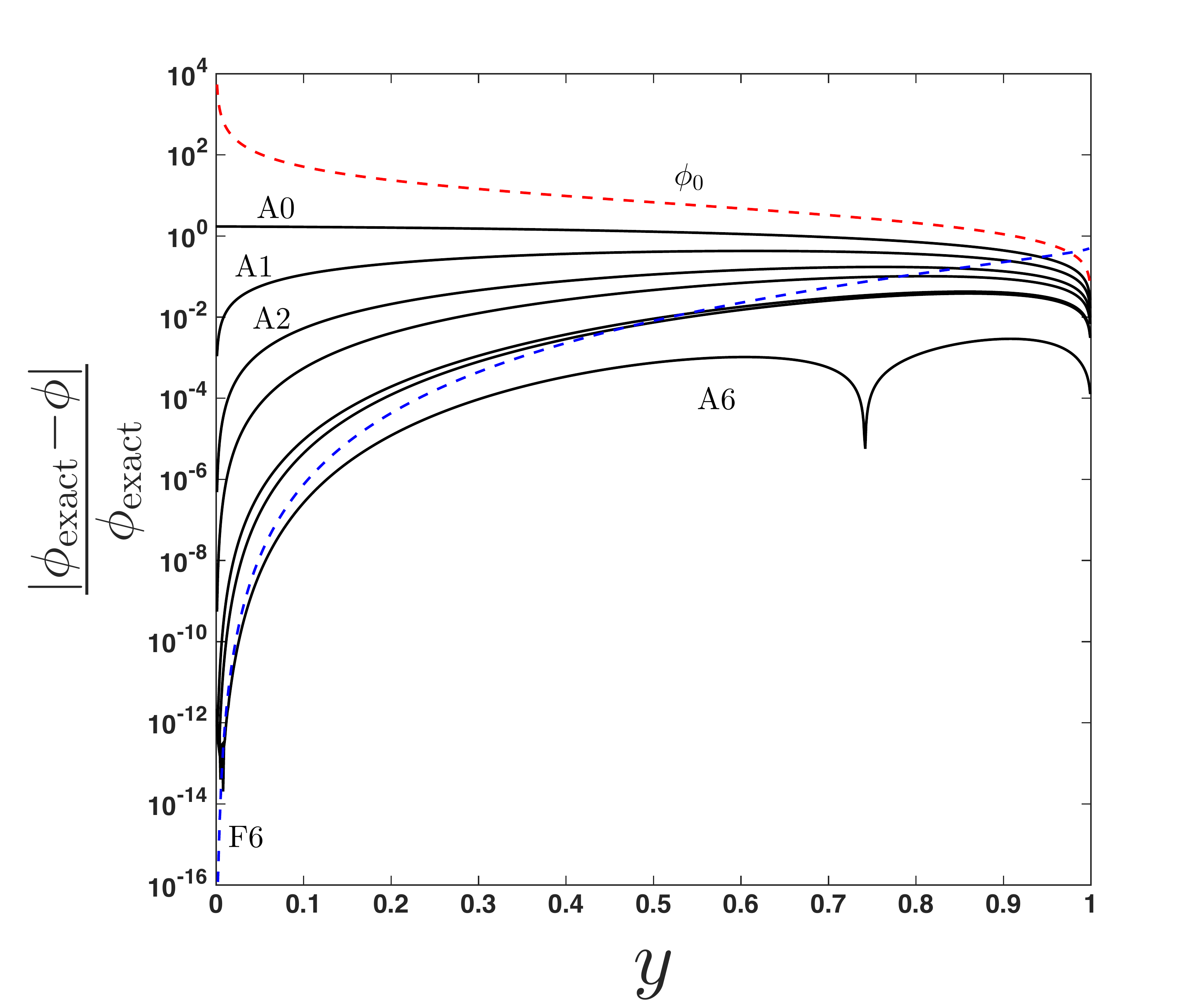}}
\subfloat{(b)\includegraphics[width=2.8in]{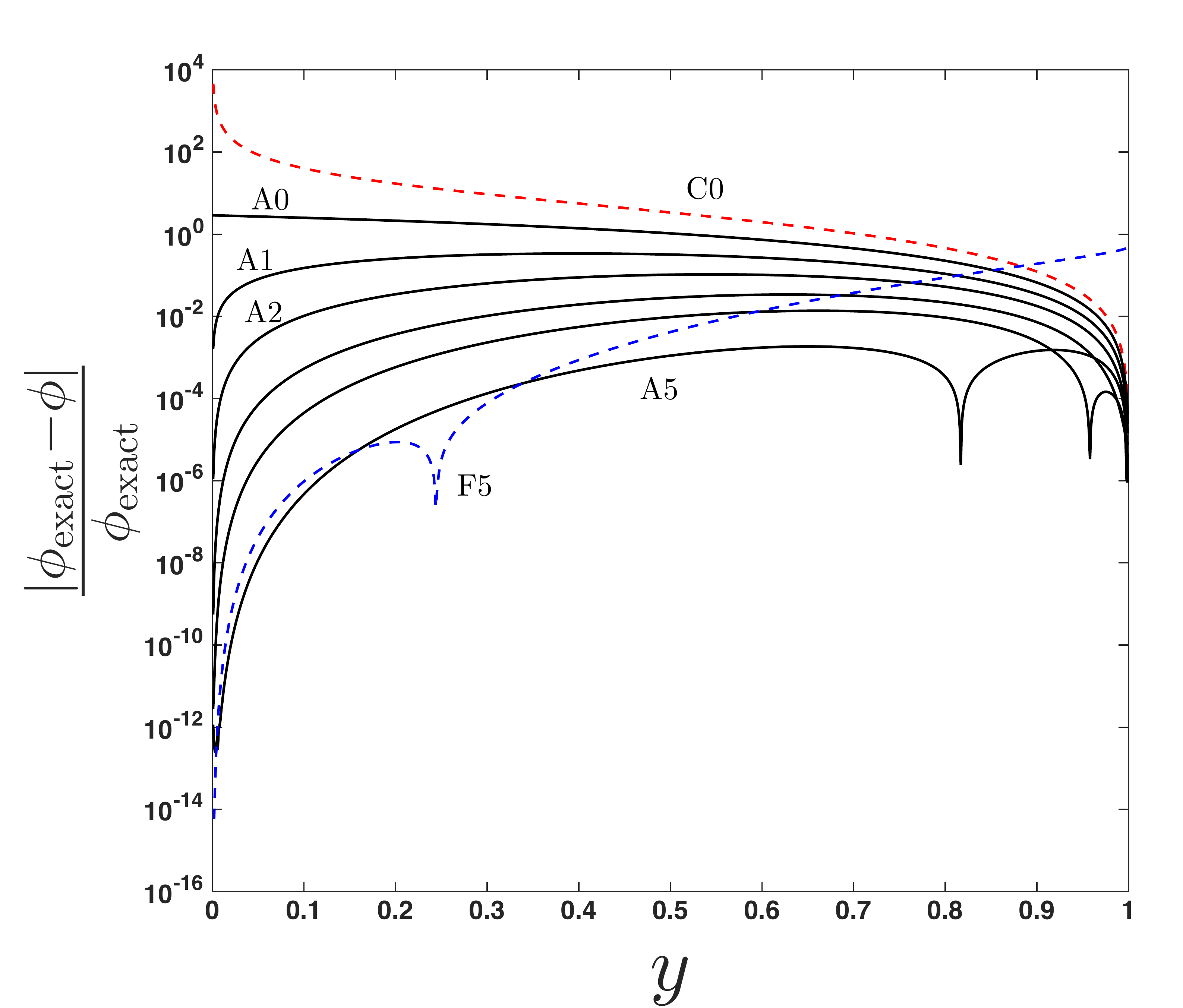}}\\
\subfloat{(c) \includegraphics[width=2.8in]{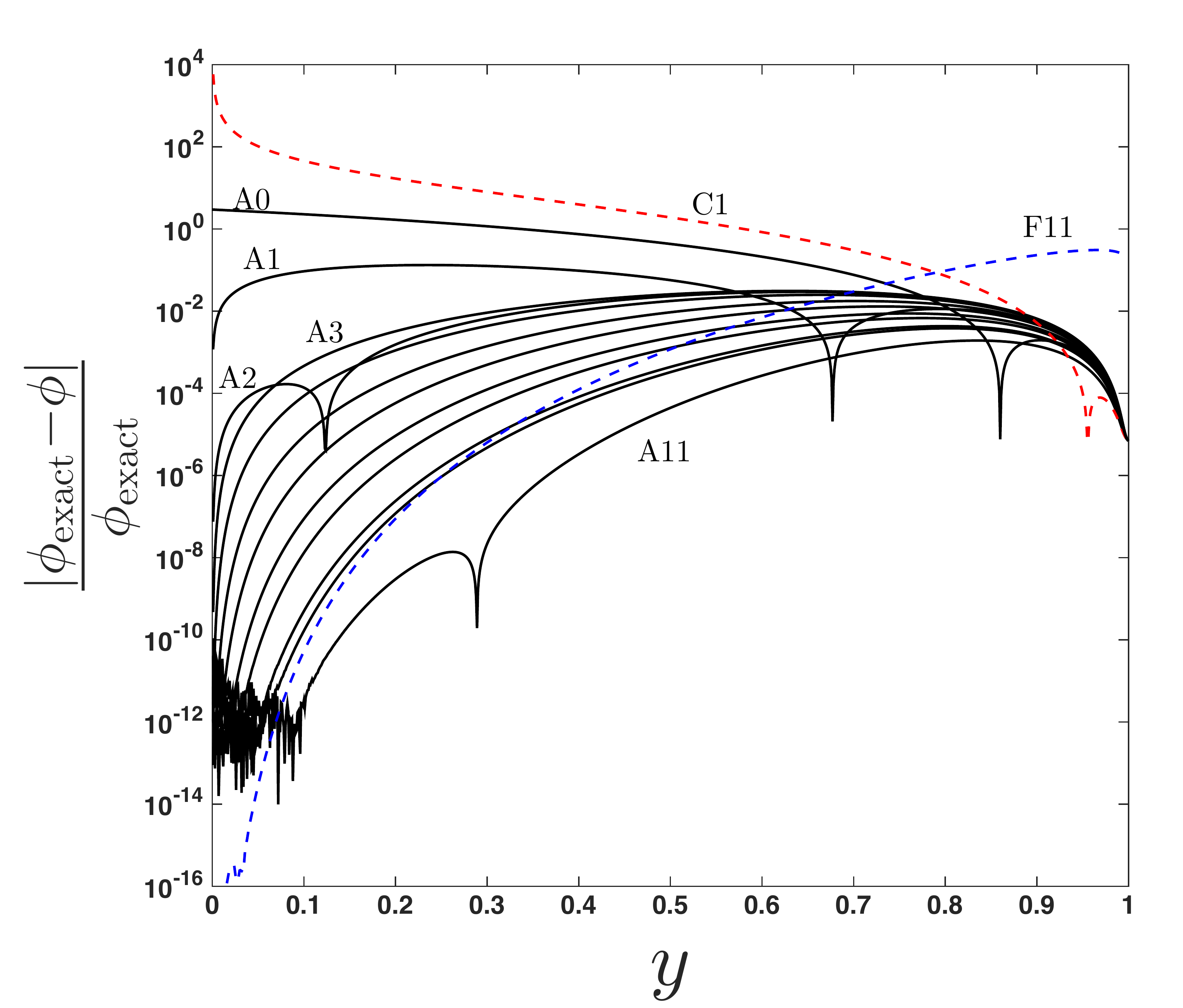}}
\subfloat{(d)\includegraphics[width=2.8in]{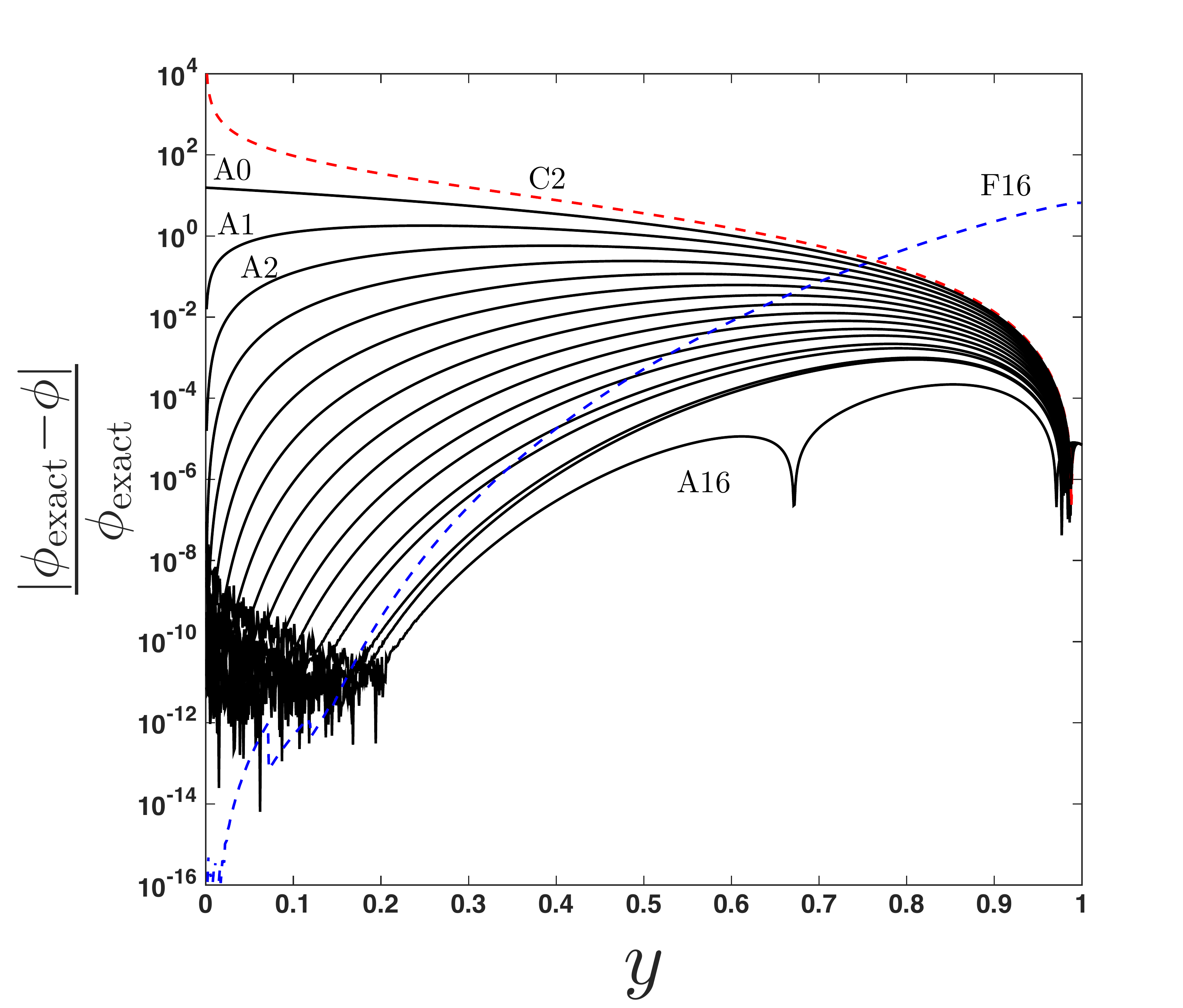}}
\end{center}
\caption{The effect of including exact higher-order terms from the closest approach expansion (CAE) (equation~(\ref{eq:CDL})) on optimal truncation of the approximant~(\ref{A}) for an extremal ($a$=1) Kerr black hole with $b'$=0.1. Relative error of the approximant A$N$ for increasing $N$ and setting (a) $K$=$-1$ (zeroth-order CAE, $\phi_0$), (b) $K$=0 (1/2-order CAE), (c) $K$=1 (3/2-order CAE), and (d) $K$=2 (5/2-order CAE) in approximant~(\ref{A}).  The error begins to increase (not shown) when carried beyond the maximum truncation $N$ shown in each plot. The far-distance series F$N$ at this final $N$ and the closest-approach series C$K$ used in the construction of the approximant are shown by dashed curves.  The cusps in the figures have no physical meaning and simply indicate where the sign of ($\phi_\mathrm{exact}-\phi$) changes.}
\label{fig:OptimalTruncation}
\end{figure*}

\begin{figure*}[h!]
\begin{center}
\subfloat{(a) \includegraphics[width=2.75in]{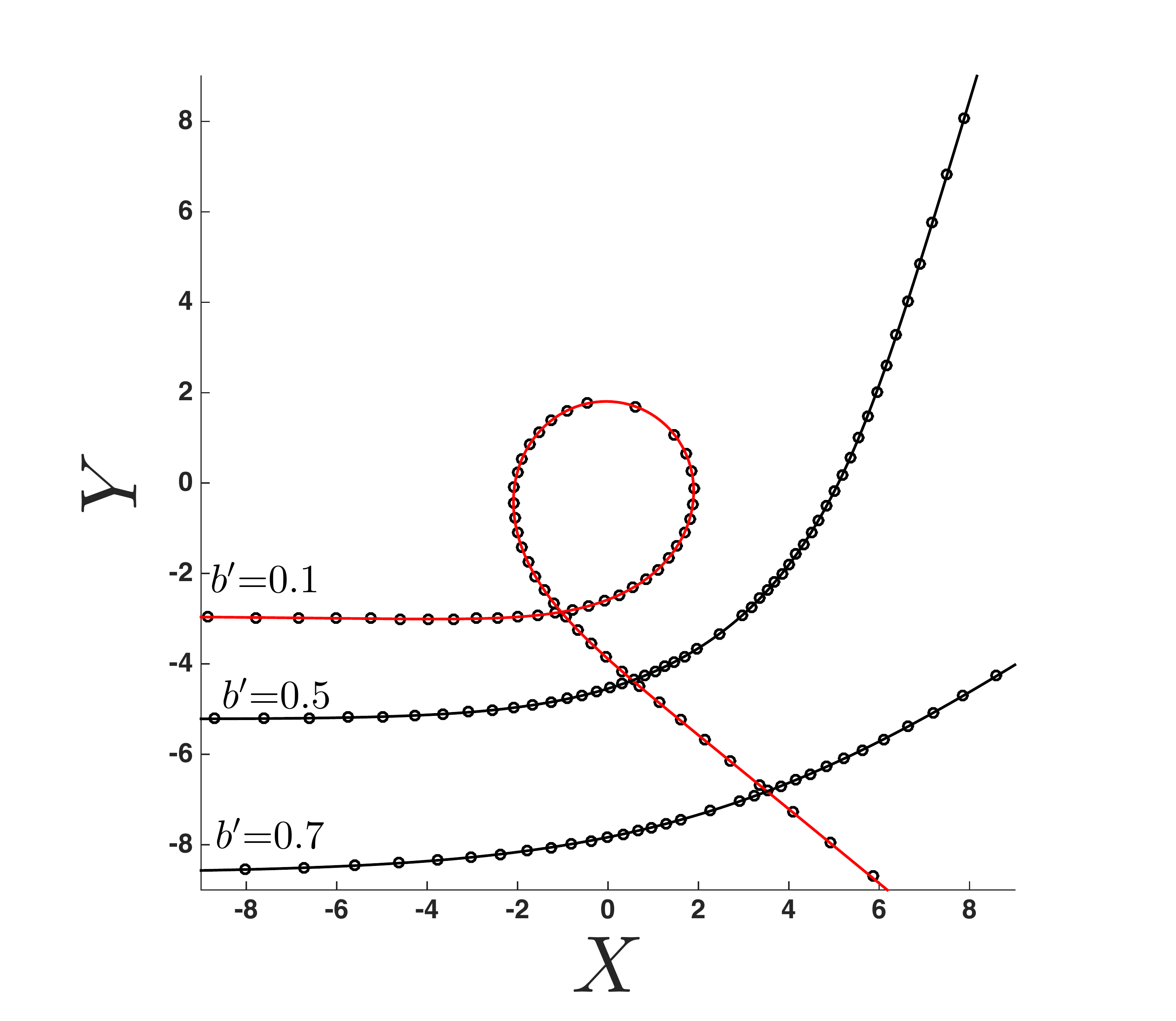}}
\subfloat{(b)\includegraphics[width=2.75in]{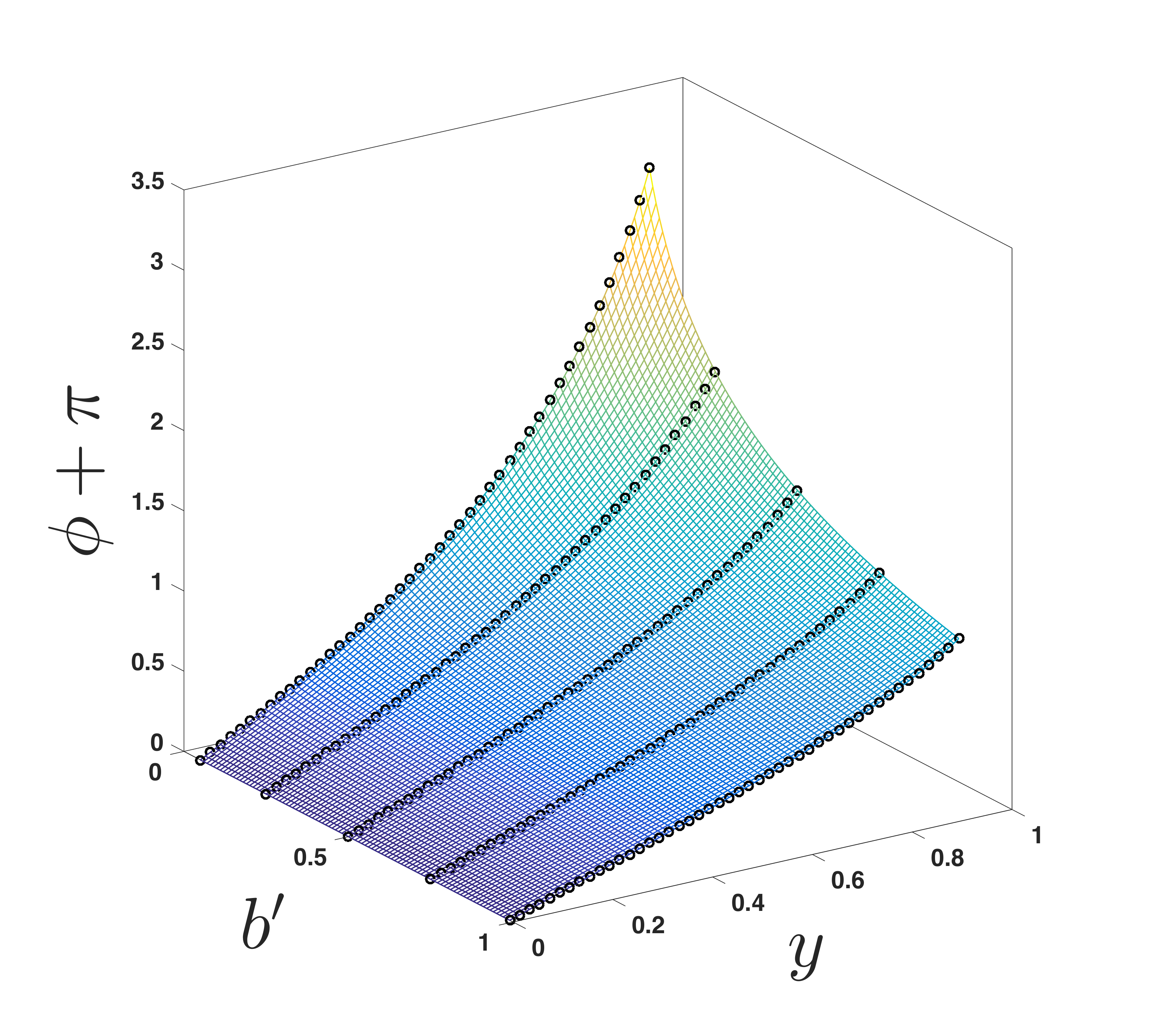}}\\
\subfloat{(c) \includegraphics[width=2.75in]{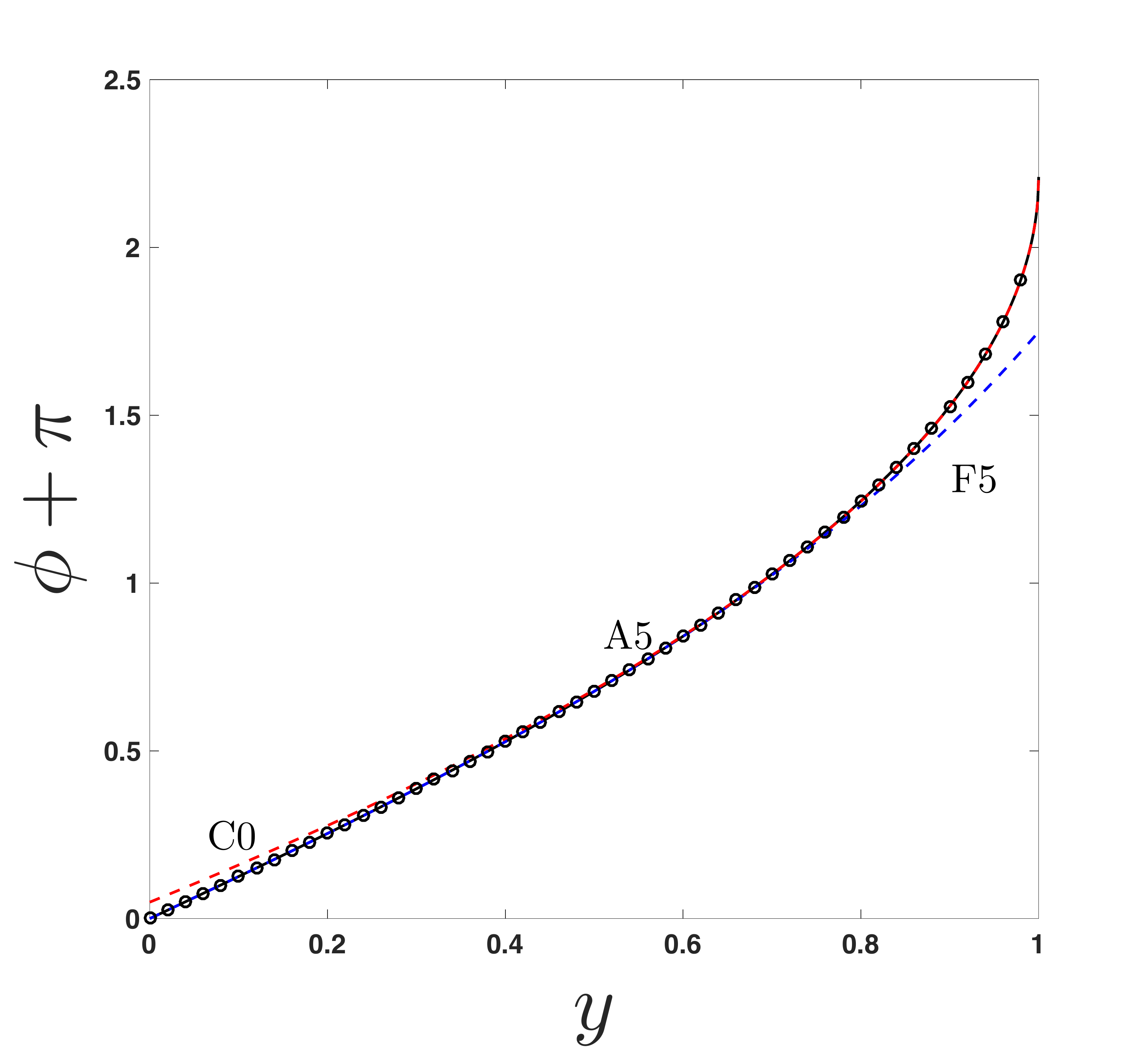}}
\subfloat{(d)\includegraphics[width=2.75in]{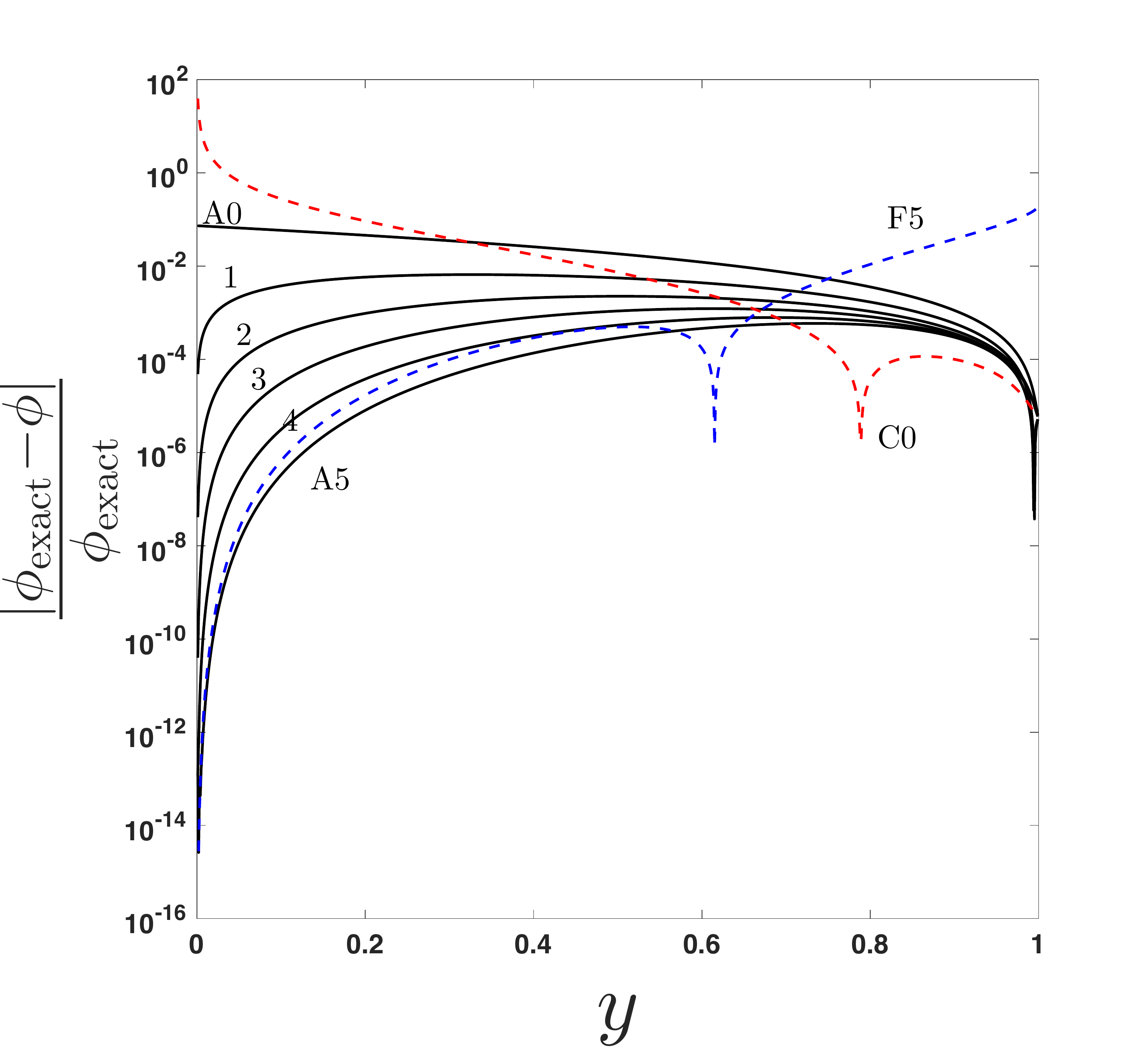}}
\end{center}
\caption{$a=0.95$: (a) Trajectory of light around a Kerr black hole, shown for different values of impact parameter $b'$.  The approximant~(\ref{A}) (solid curves) (using $N$=5, $K=-1$) are compared with the numerical solution ($\circ$) of~(\ref{eq:intY}); they are indistinguishable on the scale of the figure.  (b) parameter space for the solution of~(\ref{eq:intY}), as given by approximant~(\ref{A}) (colored mesh)  (using $N=5$, $K=0$). (c) $b'=0.5$: The approximant~(\ref{A}) (solid curve) (using $N$=5, $K=0$), the 5-term far-distance series (dashed curve, labeled F5), and the 1/2-order ($K=0$) closest-approach limit (dashed curve, labeled C0) used as an input to the approximant are compared with the numerical solution ($\circ$) of~(\ref{eq:intY}). (d) $b'=0.5$: Relative error for increasing $N$, taken up to the optimal truncation for $K=0$.   }
\label{fig:apt95}
\end{figure*}

\begin{figure*}[h!]
\begin{center}
\subfloat{(a) \includegraphics[width=2.75in]{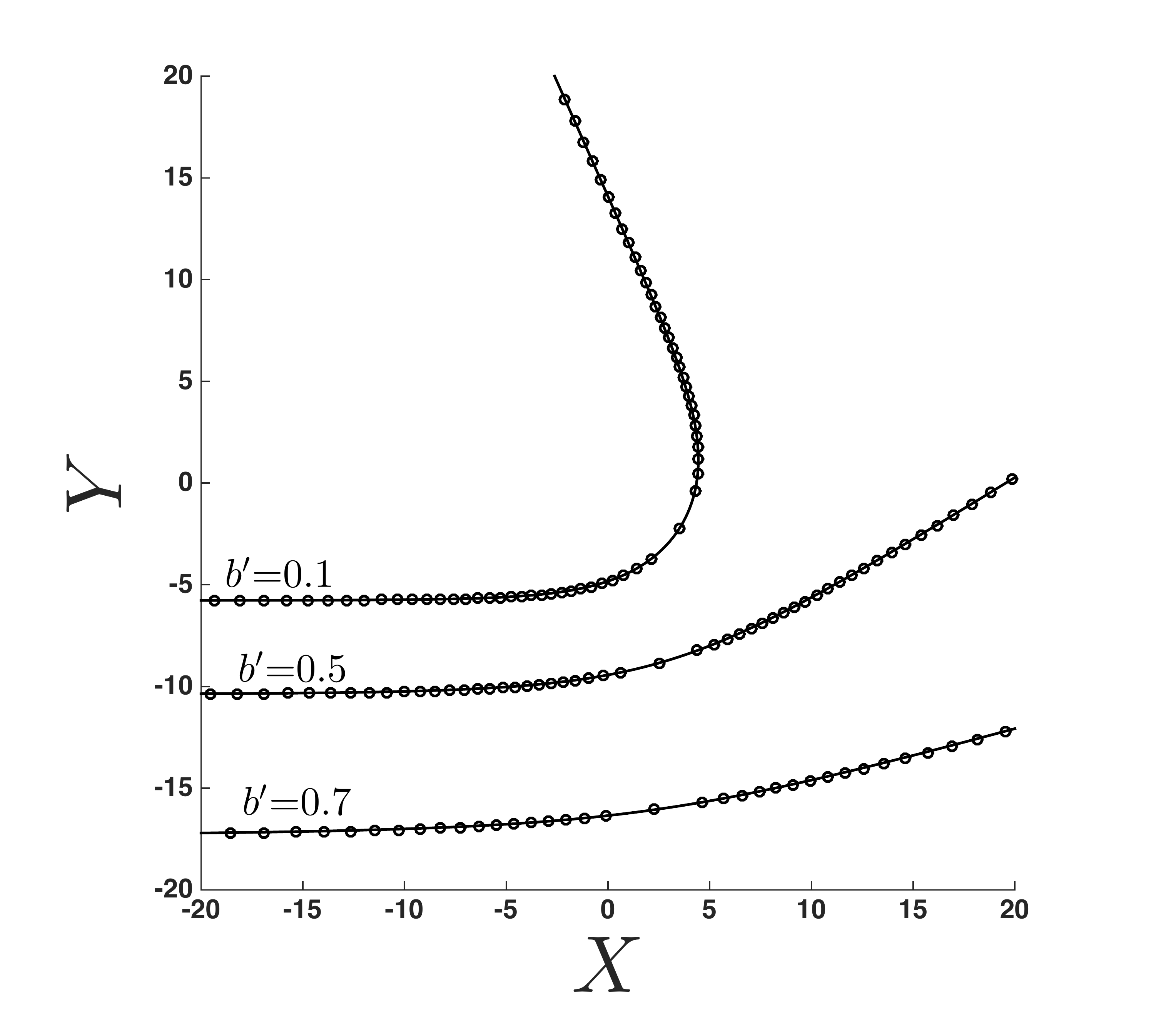}}
\subfloat{(b)\includegraphics[width=2.75in]{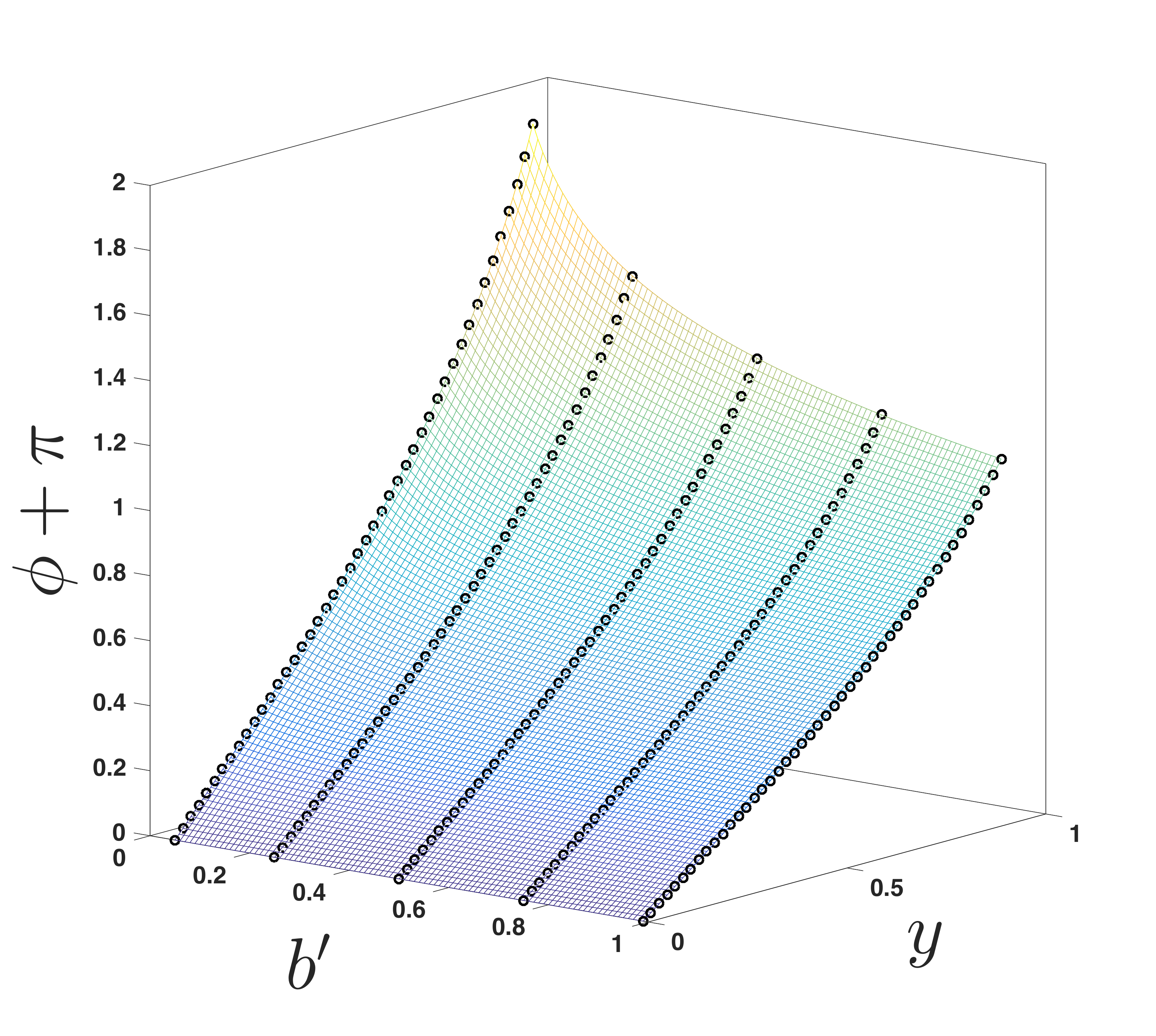}}\\
\subfloat{(c) \includegraphics[width=2.75in]{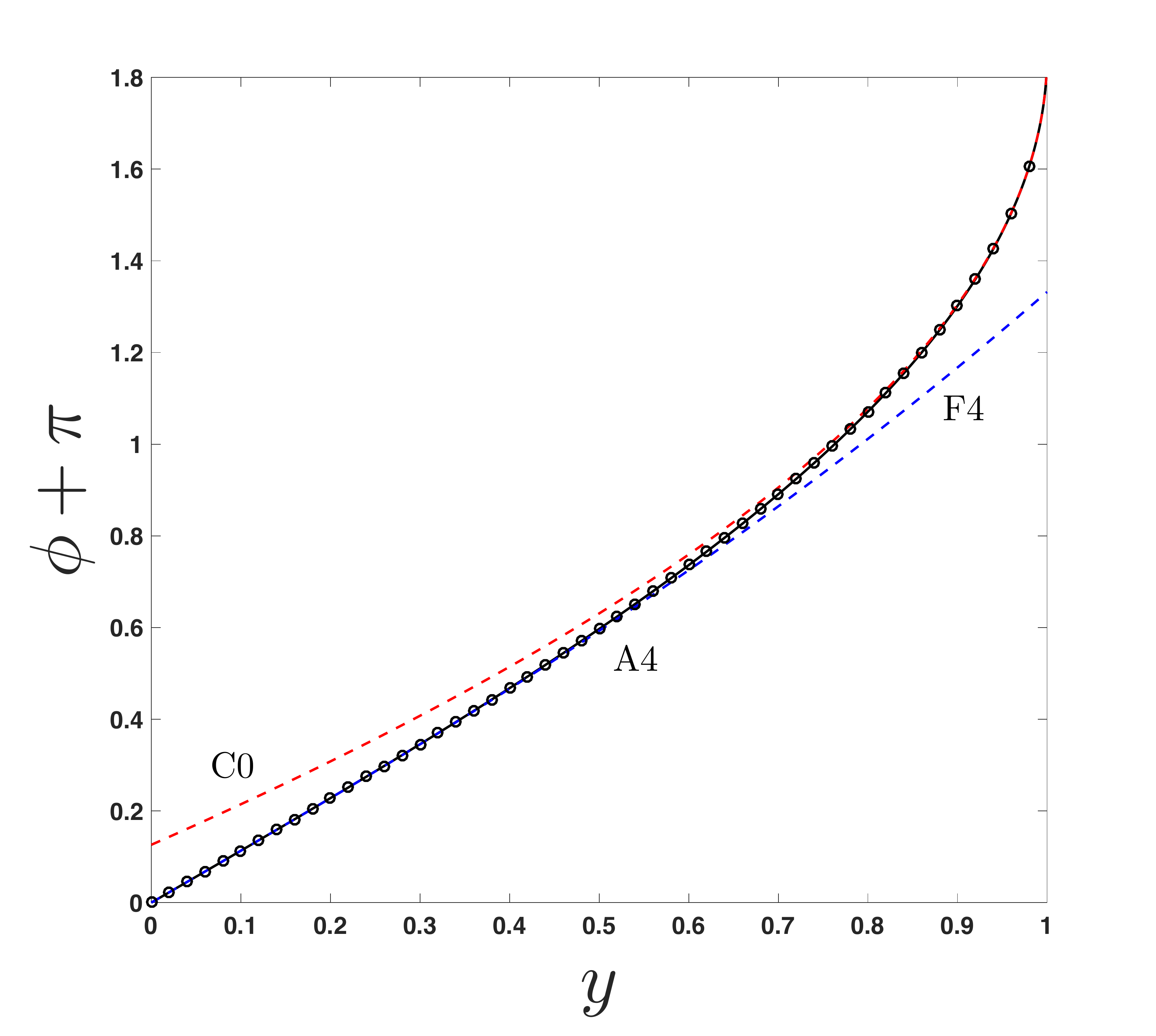}}
\subfloat{(d)\includegraphics[width=2.75in]{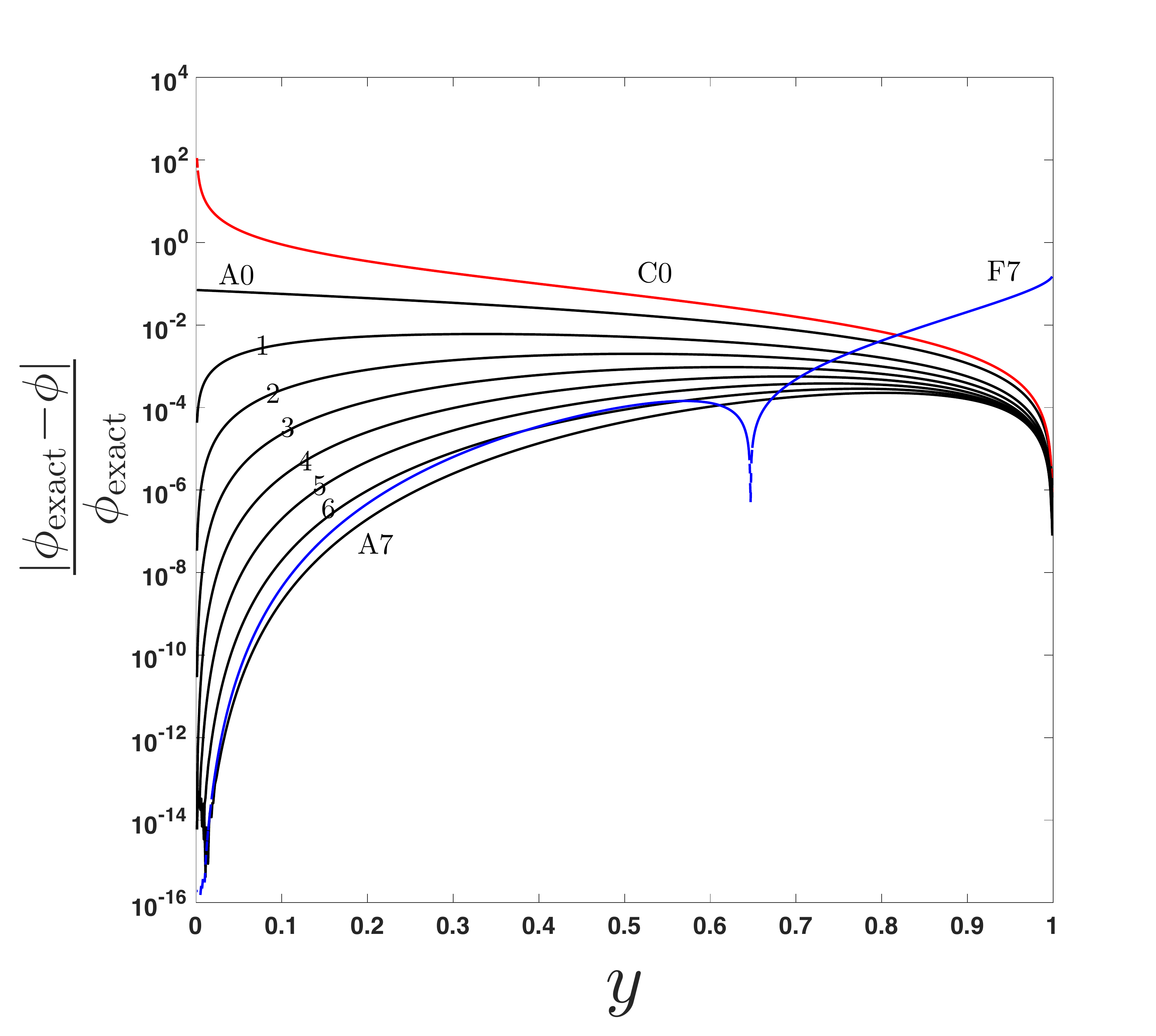}}
\end{center}
\caption{(a) Trajectory of light around a Schwarzchild ($a=0$) black hole, shown for different values of impact parameter $b'$.  The approximant~(\ref{A}) (solid curves) (using $N$=4, $K=-1$) are compared with the numerical solution ($\circ$) of~(\ref{eq:intY}); they are indistinguishable on the scale of the figure.  (b) parameter space for the solution of~(\ref{eq:intY}), as given by approximant~(\ref{A}) (colored mesh)  (using $N=7$, $K=0$). (c) $b'=0.5$: The approximant~(\ref{A}) (solid curve) (using $N$=4, $K=0$), the 4-term far-distance series (dashed curve, labeled F4), and the 1/2-order ($K=0$) closest-approach limit (dashed curve, labeled C0) used as an input to the approximant are compared with the numerical solution ($\circ$) of~(\ref{eq:intY}). (d) $b'=0.5$: Relative error for increasing $N$, taken up to the optimal truncation for $K=0$.   }
\label{fig:a0}
\end{figure*}

The approximant~(\ref{A}) is compared with the numerical solution of~(\ref{eq:intY}) in figures~\ref{fig:trajectory} through~\ref{fig:a0}.       All trajectories in the figures are constructed using either the approximant (solid curves) or the numerical solution ($\circ$'s), with the aid of the symmetry relations given in~\ref{sec:Symmetry}.   Figure~\ref{fig:trajectory} shows the trajectory of photons around an extremal ($a=1$) BH, presented in two coordinate systems: the physical $X$-$Y$ plane (see figure~\ref{fig:definition}), used in figures~\ref{fig:trajectory}a and~\ref{fig:trajectory}b; and the mapped coordinate system (see figure~\ref{fig:schematic}b), used in figures~\ref{fig:trajectory}c and~\ref{fig:trajectory}d.    In what follows we present results in both coordinate systems, as the former is physically appealing, while the latter is more sensitive to the accuracy of the approximant.  

In figure~\ref{fig:trajectory}a trajectories for $b'$= 0.1, 0.5, and 0.7 are  produced by the reduced $K=-1$ approximant given by~(\ref{approximant}) using $N$=6.  Note that, even for the $b'$=0.1 curve where photons pass thrice around the BH, the $K=-1$ approximant is indistinguishable from the numerical solution on the scale of the figure.  This is almost the case for the even closer trajectory ($b'$=0.05) shown in figure~\ref{fig:trajectory}b, where photons pass around the black hole six times.  Here, the $K=-1$ approximnant~(\ref{approximant}) (dotted line) becomes misaligned with the numerical solution (see figure inset) as the photons move radially closer to the BH.  This issue is corrected by including more information from the closest-approach limit, as shown by approximant~(\ref{A}) using $K=1$ (the solid curve of figure~\ref{fig:trajectory}b, see inset).    While the difference in approximants is difficult to see on the scale of figure~\ref{fig:trajectory}b, the mapped coordinate system (showing the same information) in figure~\ref{fig:trajectory}c clearly shows the error in the $K=-1$ approximant, visible between $y\approx0.7$ and $y\approx0.95$.   Here, the $K=-1$ approximant~(\ref{approximant}) (labeled A6) uses only information from $\phi_0$ (indicated on figure) in the closest-approach limit and the far-distance series~(\ref{eq:fdl}) using $N$=6 terms (labeled F6 on figure).  The higher-order approximant~(\ref{A}) (also shown in figure~\ref{fig:trajectory}c, labeled A15) uses information from the closest-approach expansion~(\ref{eq:CDL}) (labeled C1) taken to 3/2-order (i.e. $K=1$) and the far-distance series~(\ref{eq:fdl}) using $N$=15 terms (labeled F15 on figure); this curve is indistinguishable from the numerical solution on the scale of the figure.   This approximant is extended over a range of $b'$ to produce the surface shown in figure~\ref{fig:trajectory}d.  

Note that the value of $N$ is different  for each of the two approximants discussed above (and used in figure~\ref{fig:trajectory}). This is attributed to the fact that  the sequences of approximants we have constructed are in fact divergent past a certain order, and there is an optimal number of terms to obtain a best approximation in each case.  Although both the closest-approach and far-distance expansions are known to infinite order, they are divergent series, each having a radius of convergence (dependent on $a$ and $b'$) within the physical domain $y\in[0,1]$ as shown by the plots in figure~\ref{fig:DS_CDL} (Section~\ref{sec:closest}) and figure~\ref{fig:DS_FDL} (Section~\ref{sec:far}).  That said, the singularities responsible for the radius of convergence are non-physical, as they are not resident in the integrand of~(\ref{eq:intY}) in the interval $y\in[0,1]$.  The effect of this is that the approximant has an `optimal truncation'~\cite{Bender}, $N=N_\mathrm{opt}$, such that the $(N_\mathrm{opt}+1)^\mathrm{th}$ term leads to a smaller contribution to the approximant than both lower and higher-order terms (i.e. the remainder series diverges).   This feature is illustrated in figure~\ref{fig:extremalCauchy} for an extremal ($a=1$) BH with $b'=0.1$ using the reduced $K=-1$ approximant~(\ref{approximant}); here, $N_\mathrm{opt}=6$, as indicated in figure~\ref{fig:extremalCauchy}b.

As additional terms are included from the closest-approach series (increasing $K$) in approximant~(\ref{A}), $N_\mathrm{opt}$ changes.  The effect of $K$ on optimal truncation and achievable accuracy in the approximant is shown in figure~\ref{fig:OptimalTruncation}, where the relative error is plotted for $K=-1,~0,~1$, and 2 for an extremal ($a=1$) BH with $b'=0.1$.  In each subfigure (representing different $K$), $N$ is taken up to its optimal truncation, as discussed above. Note that for each $K$, as $N$ increases, the error decreases in \textit{some} region of $y$ (prior to optimal truncation) but the norm of the error never goes below $O$(10$^{-4}$).  In Section~\ref{sec:conclusions}, we suggest possible ways in which this persistent error may be reduced.

In order to demonstrate its versatility, the approximant is shown for $a=0.95$ and $a=0$ in figures~\ref{fig:apt95} and \ref{fig:a0}, respectively.  In each figure,  4 plots are provided: (a) the photon trajectory, using the reduced $K=-1$ approximant~(\ref{approximant}); (b) the $\phi$ surface in the $y$ vs. $b'$ plane (using, if required, the higher order approximant~(\ref{A}), i.e., $K\neq1$); (c) a cross section of the $\phi$ surface at $b'=0.5$, showing a comparison between the $N$-term far-distance (F$N$) and $K$-term closest-approach (C$K$, or $\phi_0$ for $K=-1$) expansions that are used as an input to the approximant, where $N$ is taken to be $\le N_\mathrm{opt}$ such that the approximant is accurate on the scale of the figure; and (d) the relative error of approximant~(\ref{approximant}) for increasing $N$ (up to optimal truncation $N_\mathrm{opt}$) at a fixed $K$.   The values of $a$ and $b'$ for these figures are chosen to show the range of the approximant's usage while preserving accuracy. Note that accuracy is maintained for other values of $a$ and $b'$ provided that additional terms are used from the closest-approach expansion as $b'\to0$ and additional terms are used from the weak-field limit (in $\phi_0$) as $a\to1$, as discussed in Paper 1.

We now provide some guidance on using approximant~(\ref{A}) and its reduced form~(\ref{approximant}).  For plotting trajectories of any spin $a$ in the range $b'\in[0.1, 1]$, such as those shown in figures~\ref{fig:trajectory}a,~\ref{fig:trajectory}b,~\ref{fig:apt95}a, and~\ref{fig:a0}a, one may use the reduced $K=-1$ approximant~(\ref{approximant}) for simplicity's sake.  If used for visualization purposes, these trajectories will be indistinguishable from the numerical solution, with an error of (at most) $O$($10^{-3}$).  For closer trajectories ($b'<0.1$), or in any case where one wishes to reduce local error, we recommend using the more general approximant~(\ref{A}), which includes corrections from the closest-approach expansion. One may then determine the optimal truncation by following the procedure illustrated in figure~\ref{fig:extremalCauchy}b, which does not require knowledge of the exact solution.  

In regards to computational expense, the authors have found the closed-form solution given by~(\ref{A}) to be, in general, an order of magnitude faster than the numerical evaluation\footnote{The numerical evaluation of~(\ref{eq:intY}) is done using the ``quad'' command in MATLAB, which uses an adaptive recursive Simpson's rule~\cite{Gander}} of~(\ref{eq:intY}), when generating the same number of data points.  Furthermore, the authors found that time-savings from using less terms (either in $K$ or $N$) in the approximant are negligible.  We recommend using as many terms as are required for the desired accuracy, as it will be faster than solving the exact elliptic integrals numerically.

\section{Conclusions}\label{sec:conclusions}
Analytical expressions that describe the full trajectory of photons propagating in the equatorial plane of a Kerr black hole are obtained using asymptotic approximants.  These have potential use in future ray-tracing efforts and radiation transport numerical projects, particularly those studying accretions disks around black holes and their observable electromagnetic emission.   The expressions obtained provide accurate trajectory predictions for arbitrary spin and impact parameters, and provide significant time advantages compared with numerical evaluation of the elliptic integrals that describe photon trajectories.  The asymptotic approximants provided here are accurate closed-form expressions that bridge the weak-field (large impact parameter), strong-field (near-critical impact parameter), far-distance (large radius), and closest approach (smallest radial distance from the black hole) limits.  To that end, asymptotic expansions are derived for the azimuthal angle in the far-distance and closest-approach  limits, and new coefficients are reported for the bending angle in the weak-field limit.

While an optimal truncation of the approximants provided here can be determined to maintain accuracy, future work may focus on incorporating the non-physical singularities responsible for the divergence of the far-distance and closest approach series.  One idea would be to use a modified Pad\'e approach to incorporate these singularities as poles, while maintaining the correct asymptotic behavior in all regions listed above.  This would involve a matrix inversion in the process of computing the approximant coefficients, whereas now their computation is solely recursive.  That said, there are efficient methods for computing this particular inversion~\cite{Trefethen} and this would still likely require less computational time than the full numerical solution of the elliptic integrals.  

We close this paper by distinguishing the approach taken here, i.e., asymptotic approximants, from that of asymptotic matching used in singular perturbation theory.  In the later approach, a region of overlap must occur in which two asymptotic expansions are valid, and a systematic method is used to obtain an expression that is uniformly valid over the whole domain.   The key is that, in asymptotic matching, the physical domain is completely encompassed and described by overlapping asymptotic expansions.  Although asymptotic approximants do connect two asymptotic expansions in different regions of a domain, it is not necessary that the original asymptotic expansions overlap, and there can be a gap in which both expansions are not valid (see, for example, figure~\ref{fig:divergence}).  By forming a convergent sequence (via analytic continuation) of approximants, the gap region is well approximated where both asymptotic series can fail.

 \ack{The authors wish to thank Y. Zlochower for helpful conversations, as well as the critical assistance of the RIT Department of Access Services, including P. Arndt, L. Braggiotti, B. DeGroote, H. Jentsch, L. Joslyn, D. Moore, M. Murphy, C. Reminder, and J. Riccardi, among many others.  JAF was supported by NSF award  ACI-1550436.  RJB and JAF were supported by NSF award PHY-1659740.}

\section*{References}
\bibliography{Approximants}

\appendix
\section{On the Symmetry of Photon Trajectories \label{sec:Symmetry}}
\begin{figure}[h!]
\begin{center}
\includegraphics[width=6in]{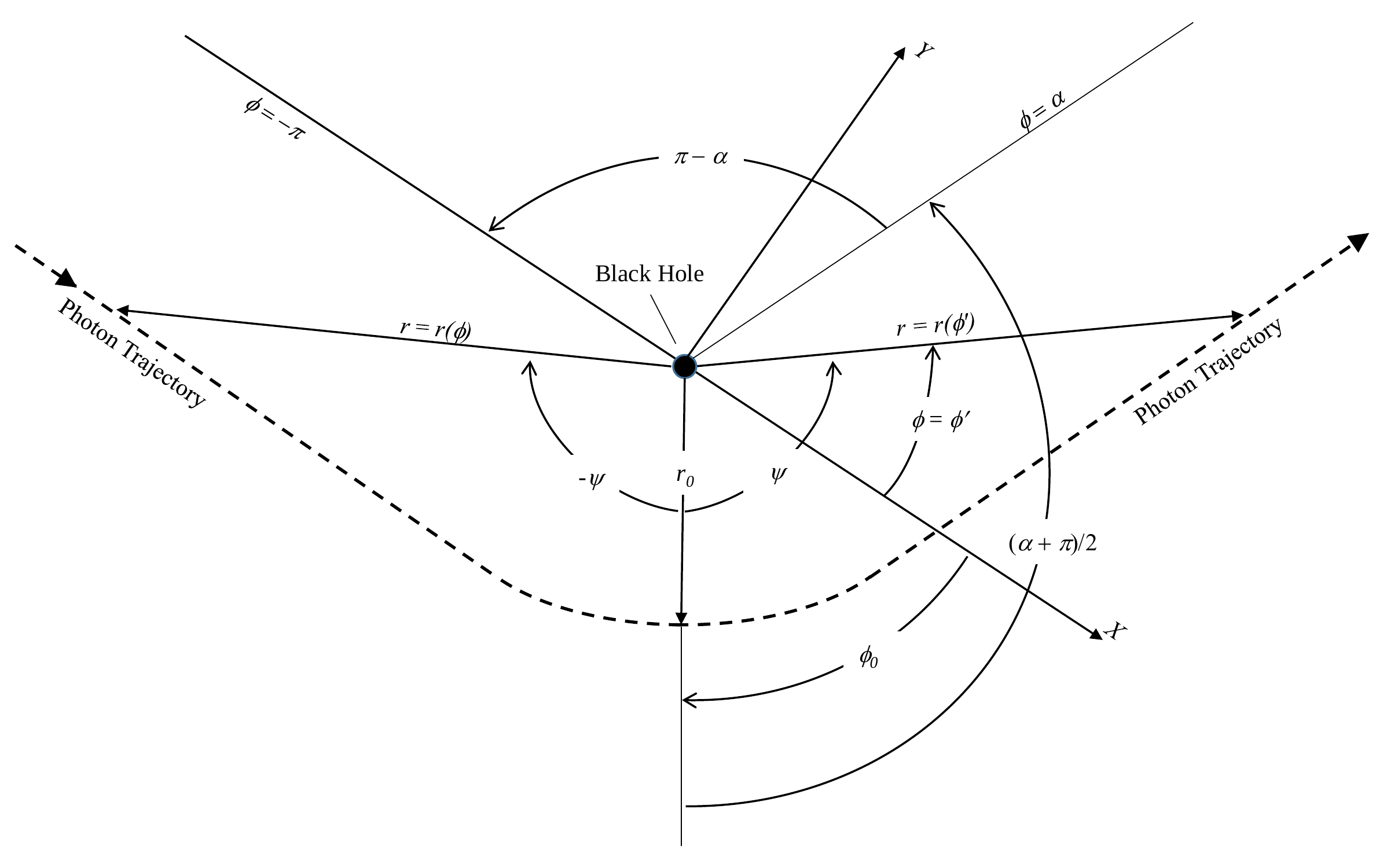}
\end{center}
\caption{Definition sketch showing the symmetry of the photon trajectory about the about the angle $\phi_0$.}
\label{fig:symmetry}
\end{figure}

It is well known that the trajectory of a photon, $r=r(\phi)$, is symmetric about the angle of closest approach to a black hole, $\phi_0$, where these quantities are defined in figure~\ref{fig:symmetry} (note that figure~\ref{fig:symmetry} and its notation is consistent with a rotated portion of the trajectory sketch of figure~\ref{fig:definition}).  Thus, it is only necessary to determine the parameterization $r=r(\phi)$ for $\phi\in[-\pi,\phi_0]$, and this can be reflected to determine the trajectory for $\phi=\phi'\in[\phi_0,\alpha]$. Here, we have introduced the angle $\phi'$ for notational convenience in what follows. We also define a more natural symmetry angle  $\psi$=$\psi(\phi)$ where $r(\phi(\psi))=\bar{r}({\psi})$ and $\bar{r}(\psi)=\bar{r}(-\psi)$, as shown in figure~\ref{fig:symmetry}.  The goal here is to express this symmetry condition in terms of the angle $\phi$, for plotting purposes. 

Noting that the symmetry angle satisfies $\phi_0<0$ in figure~\ref{fig:symmetry}, the relationship between $\psi$ and $\phi$ may be expressed as 
\begin{equation}
\psi=\phi'-\phi_0,
\label{eq:A1}
\end{equation}
and for a given $\phi\in[-\pi,\phi_0]$, there is a corresponding value of $-\psi$ (see figure~\ref{fig:symmetry}) expressed as 
\begin{equation}
-\psi=\phi-\phi_0.
\label{eq:A2}
\end{equation}
Eqs.~(\ref{eq:A1}) and~(\ref{eq:A1}) may be combined to eliminate $\psi$, leading to
\begin{equation}
\phi'=2\phi_0-\phi.
\label{eq:A3}
\end{equation}
Eq.~(\ref{eq:A3}) provides a relationship between $\phi\in[-\pi,\phi_0]$ and $\phi'\in[\phi_0,\alpha]$ that preserves the symmetry relationship. The radial parameterization that expresses symmetry can thus be expressed conveniently as:
\begin{equation}
r(2\phi_0-\phi)=r(\phi),~\phi\in[-\pi,\phi_0].
\label{eq:A4}\end{equation}
Finally, the geometry of~\ref{fig:symmetry} indicates that $\phi_0$ is related to $\alpha$, the bending angle shown in figure~\ref{fig:schematic}, as $\alpha-\phi_0=(\alpha+\pi)/2$, or 
\begin{equation}
\phi_0=(\alpha-\pi)/2.
\label{eq:AA5}
\end{equation}
Eqs.~(\ref{eq:A4}) and~(\ref{eq:AA5}) provide the desired symmetry description, and enable the full photon trajectory to be determined by evaluating $r(\phi)$ for $\phi\in[-\pi,\phi_0]$.  Note that in practice, $\phi$ is our dependent variable, and so we choose a suitable $r$ range to canvas $\phi\in[-\pi,\phi_0]$, evaluate $\phi(r)$ either numerically or using approximant~(\ref{A}), compute $\phi'$ using~(\ref{eq:A3}), and manually piece $\phi$ and $\phi'$ together on the same plot (see the figures in Section~\ref{sec:Results}).

\section{Recursive Formulation of Coefficients \label{sec:Recursion}} 
The following relations are used to develop recursive formulae for the coefficients of~(\ref{eq:fdl}),~(\ref{eq:CDL}), and~(\ref{A}) (and thus its simplification~(\ref{approximant})).  The first relation is the well-known Cauchy product of two series~\cite{Churchill}:
\begin{equation}
\sum_{n=0}^\infty a_ny^n\sum_{n=0}^\infty b_ny^n=\sum_{n=0}^\infty\left(\sum_{j=0}^na_jb_{n-j}\right)y^n. 
\label{eq:Cauchy}
\end{equation}
If one sets the left-hand side of~(\ref{eq:Cauchy}) equal to unity, this may be re-arranged to represent the expansion of the recipricol of a series.   The coefficients of this expansion are then obtained by setting the right-hand side of~(\ref{eq:Cauchy}) equal to unity and evaluating like-powers of $y$ on each side.   This leads to a recursive representation for the expansion of the reciprocal of a series:
\begin{eqnarray}
\nonumber
\left(\sum_{n=0}^\infty a_ny^n\right)^{-1}=\sum_{n=0}^\infty b_ny^n,\\
b_{n>0}=-\frac{1}{a_0}\sum_{j=1}^na_jb_{n-j},~b_0=\frac{1}{a_0}.
\label{eq:inverse}
\end{eqnarray}
The generalization of~(\ref{eq:inverse}) for the expansion of a series raised to any real power $\nu$ is given by J. C. P. Miller's formula~\cite{Henrici}:
\begin{eqnarray}
\nonumber
\left(\sum_{n=0}^\infty a_ny^n\right)^{\nu}=\sum_{n=0}^\infty b_ny^n,\\
b_{n>0}=\frac{1}{n~a_0}\sum_{j=1}^n(j\nu-n+j)a_jb_{n-j},~b_0=(a_0)^\nu.
\label{eq:Miller}
\end{eqnarray}

To form the approximants in this paper, on the domain $y\in[0,1]$, it is useful to have an explicit formula that relates a truncated series in $y$ as a truncated series in $(y-1)$ and vice-versa.  By solving the system of linear equations required of this conversion, one arrives at the following:
\begin{eqnarray}
\nonumber 
\sum_{n=0}^Na_ny^n=\sum_{n=0}^Nb_n (y-1)^n,\\
\nonumber
a_n=\frac{1}{n!}\sum_{m=0}^N\frac{\Gamma(m+1)}{\Gamma(m-n+1)}b_m,\\
b_n=\frac{1}{n!}\sum_{m=0}^N\frac{\Gamma(m+1)(-1)^{m-n}}{\Gamma(m-n+1)}a_m,
\label{eq:conversion}
\end{eqnarray}
where one would pick the appropriate relation in~(\ref{eq:conversion}) to determine $b_n$ from $a_n$ or vice-versa.  The gamma function $\Gamma$ used in~(\ref{eq:conversion}) is a compact way to write the products that arise in solving the linear system, and is convenient to use when writing code. 

\subsection{Closest-approach series \label{sec:Recursion:Closest}}
In order to obtain the closest-approach expansion~(\ref{eq:CDL}) as $y\to1$, we Taylor expand $\mathcal{G}(z;a,b)$ (the integrand of~(\ref{eq:CDLtransformed})) about $z^2=0$ (since $z$ only appears as $z^2$ in $\mathcal{G}$); the integration (with respect to $z$) is then trivial.   Since a recursion or pattern for the expansion of $\mathcal{G}(z;a,b)$  is not immediately known, we separate it into the product of functions whose expansions about $z^2=0$ may be obtained individually. First, we decompose the function  $\mathcal{G}(z;a,b)$  (given in~(\ref{eq:CDLtransformed})) as
\begin{eqnarray*}
\mathcal{G}(z;a,b)=\left[2u_0b-4u_0^2 (b-a)(1-z^2)~\right]&\left[1-2u_0(1-z^2)+a^2u_0^2(1-z^2)^2~\right]^{-1}\\&\times\left[1+(1-z^2)-2\left(b-a\right)^2u_0^3 (1-z^2)^2~\right]^{-1/2}.
 \label{eq:G}
 \end{eqnarray*}
 In preparation for the use of identities~(\ref{eq:inverse}) and~(\ref{eq:Miller}), it is notationally convenient to next rewrite the above as
 \begin{eqnarray*}
\mathcal{G}(z;a,b)=\left[\sum_{n=0}^\infty P_nz^{2n}\right]\left[\sum_{n=0}^\infty s_nz^{2n}\right]^{-1}\left[\sum_{n=0}^\infty q_nz^{2n}\right]^{-1/2},
 \label{eq:G2}
 \end{eqnarray*}
 where $P_0=2u_0b-4u_o^2(b-a),~P_1=4u_0^2(b-a),~s_0=1+u_0(-2+a^2u_0),~s_1=2u_0-2a^2u_0^2,~s_2=a^2u_0^2,~q_0=2[1-u_0^3(b^2-2ab+a^2)],~q_1=4u_0^3(a-b)^2-1,~q_2=-2u_0^3(b-a)^2$, and $P_{n\ge2}=s_{n\ge3}=q_{n\ge3}=0$.  The second and third bracketed items above are now expanded respectively using the recursive forms~(\ref{eq:inverse}) and~(\ref{eq:Miller}) (letting $\nu=-1/2$), leading to
\begin{eqnarray}
\mathcal{G}(z;a,b)=\sum_{n=0}^\infty P_nz^{2n}\sum_{n=0}^\infty S_nz^{2n}\sum_{n=0}^\infty Q_nz^{2n},
 \label{eq:G3}
 \end{eqnarray}
where 
\[S_{n>0} =-\frac{1}{s_0}\sum_{j=1}^{n}s_jS_{n-j},~S_0=1/s_0,\]
and
\[Q_{n>0}=\frac{1}{nq_0}\sum_{j=1}^{n}(\frac{j}{2}-n)q_jQ_{n-j},~Q_0=1/\sqrt{q_0}.\]
Finally, Cauchy's product rule~(\ref{eq:Cauchy}) is applied twice to~(\ref{eq:G3}) to obtain 
\begin{eqnarray*}
\mathcal{G}(z;a,b)=\sum_{n=0}^\infty \tilde{C}_n\hat{y}^{n},
 \label{eq:G4}
 \end{eqnarray*}
where 
\[\tilde{C}_n=\sum_{k=0}^{n} \left(\sum_{j=0}^{k} P_j S_{k-j} \right) Q_{n-k}.\]

\subsection{Far-distance series \label{sec:Recursion:Far}}
In order to obtain the far-distance expansion~(\ref{eq:fdl}) as $y\to0$, we firstTaylor expand $g(\hat{y};a,b)$ (the integrand of~(\ref{eq:intY})) about $\hat{y}=0$; the integration (with respect to $\hat{y}$) is then trivial.   Since a recursion or pattern for the expansion of $g(\hat{y};a,b)$ is not immediately known, we separate it into the product of functions whose expansions about $y=0$ may be obtained individually. First, we decompose the function $g(\hat{y};a,b)$ (given in~(\ref{eq:intY})) as
\begin{eqnarray*}
 g(\hat{y};a,b)=\left[u_0b-2u_0^2 (b-a)\hat{y}\right]\left[1-2u_0\hat{y}+a^2u_0^2\hat{y}^2\right]^{-1}\left[2(b-a)^2u_0^3\hat{y}^3-(b^2-a^2)u_0^2\hat{y}^2+1\right]^{-1/2}.
 \label{eq:g}
 \end{eqnarray*}
 In preparation for the use of identities~\ref{eq:inverse}) and~(\ref{eq:Miller}), it is notationally convenient to next rewrite the above as
 \begin{eqnarray*}
g(\hat{y};a,b)=\left[\sum_{n=0}^\infty p_ny^n\right]\left[\sum_{n=0}^\infty d_ny^n\right]^{-1}\left[\sum_{n=0}^\infty c_ny^n\right]^{-1/2},
 \label{eq:g2}
 \end{eqnarray*}
 where $p_0=u_0b,~p_1=-2u_0^2 (b-a),~p_{k>1}=0,~d_0=1,~d_1=-2u_0,~d_2=a^2u_0^2,~d_{k>2}=0,~c_0=1,~c_1=0,~c_2=-(b^2-a^2)u_0^2,~c_3=2(b-a)^2u_0^3$, and $c_{k>3}=0$.  The second and third bracketed items above are now expanded respectively using the recursive forms~(\ref{eq:inverse}) and~(\ref{eq:Miller}) (letting $\nu=-1/2$), leading to
\begin{eqnarray}
 g(\hat{y};a,b)=\sum_{n=0}^\infty p_ny^n\sum_{n=0}^\infty F_ny^n\sum_{n=0}^\infty \tilde{c}_ny^n,
 \label{eq:g3}
 \end{eqnarray}
where 
\[F_{n>0}=-\sum_{k=1}^nd_kF_{n-k},~F_0=1,\]
and
\[\tilde{c}_{n>0}=\frac{1}{n}\sum_{k=1}^n\left(\frac{k}{2}-n\right)c_k\tilde{c}_{n-k},~\tilde{c}_0=1.\]
Finally, Cauchy's product rule~(\ref{eq:Cauchy}) is applied twice to~(\ref{eq:g3}) to obtain 
\begin{eqnarray*}
 g(\hat{y};a,b)=\sum_{n=0}^\infty g_n\hat{y}^{n},
 \label{eq:g4}
 \end{eqnarray*}
where 
\[g_n=\sum_{j=0}^{n} \left(\sum_{k=0}^{j} p_k F_{j-k} \right) \tilde{c}_{n-j}.\]

\subsection{Coefficients of the asymptotic approximant \label{sec:Recursion:Approximant}}
In order to obtain the approximant coefficients $A_n$ given in~(\ref{A}) (and its reduced $K=-1$ form~(\ref{approximant})), we start with the requirement that the infinite-term expansion of the approximant~(\ref{A}) about $y=0$ be exactly equal to the infinite-term (and exactly known) far-distance series~(\ref{eq:fdl}) (i.e. replacing the left-hand side of~(\ref{A}) with the right-hand side of~(\ref{eq:fdl})):
\begin{eqnarray*}
\sum_{n=0}^\infty \tilde{g}_ny^{n}=\phi_0+\sqrt{1-y}\left\{\left[\sum_{n=0}^K C_n(y-1)^n\right]+(y-1)^{K+1}\sum_{n=0}^\infty A_n(y-1)^n\right\}.
\label{eq:step1}
 \end{eqnarray*}
Next, we solve for the $A_n$ series:
\begin{eqnarray*}
\sum_{n=0}^\infty A_n(y-1)^n=\left\{\left[-\phi_0+\sum_{n=0}^\infty\tilde{g}_n y^n\right]\left[(1-y)^{-1/2}\right]-\left[\sum_{n=0}^K C_n(y-1)^n\right]\right\}\left[(y-1)^{-K-1}\right].
\end{eqnarray*} 
To prepare for Cauchy's product rule, we next represent all $\left[~\right]$ bracketed items above as their infinite-term Taylor expansions about $y=0$, making use of~(\ref{eq:conversion}) to transform the series in $(y-1)$ into a series in $y$ as follows,
 \begin{eqnarray*}
\sum_{n=0}^\infty &A_n(y-1)^n=\nonumber \\ &\left\{\left[\sum_{n=0}^\infty\tilde{\tilde{g}}_n y^n\right]\left[\frac{1}{\sqrt{\pi}}\sum_{n=0}^\infty \frac{\Gamma(n+\frac{1}{2})}{\Gamma(n+1)}y^n\right]-\left[\sum_{n=0}^\infty\left(\frac{1}{n!}\sum_{j=0}^K \frac{(-1)^{j-n}\Gamma(j+1)}{\Gamma(j-n+1)}C_j\right)y^n\right]\right\}\nonumber \\
&\times\left[\sum_{n=0}^\infty \frac{(-1)^{-K-1}\Gamma(K+n+1)}{n!\Gamma(K+1)}y^n\right],
\end{eqnarray*} 
where $\tilde{\tilde{g}}_0=\tilde{g}_0-\phi_0$ and $\tilde{\tilde{g}}_{n>0}=\tilde{g}_{n>0}$.  Cauchy's product rule~(\ref{eq:Cauchy}) is then applied to the first two $\left[~\right]$ bracketed items above, allowing the subtraction within the $\{~\}$ braces to be combined under one series, which is multiplied by the final bracketed item as follows,
 \begin{eqnarray*}
\sum_{n=0}^\infty A_n(y-1)^n=\left\{\sum_{n=0}^\infty T_ny^n\right\}\left[\sum_{n=0}^\infty \frac{(-1)^{-K-1}\Gamma(K+n+1)}{n!\Gamma(K+1)}y^n\right],
\end{eqnarray*} 
where the coefficients $T_n$ are given in~(\ref{A}). Another application of Cauchy's product rule~(\ref{eq:Cauchy}) to the above leads to
 \begin{eqnarray*}
\sum_{n=0}^\infty A_n(y-1)^n=\sum_{n=0}^\infty\left[\sum_{j=0}^n T_{n-j} \frac{(-1)^{-K-1}\Gamma(K+j+1)}{j!\Gamma(K+1)}\right]y^n,
\end{eqnarray*} 
which now may be truncated to any order $N$ on both sides:
 \begin{eqnarray*}
\sum_{n=0}^N A_n(y-1)^n=\sum_{n=0}^N\left[\sum_{j=0}^n T_{n-j} \frac{(-1)^{-K-1}\Gamma(K+j+1)}{j!\Gamma(K+1)}\right]y^n.
\end{eqnarray*} 
Finally, the $A_n$ coefficients (as written in~(\ref{A})) are found using the formulae given in~(\ref{eq:conversion}) to convert between series in $y$ and series in $(y-1)$. 

\section{5$^{th}$-order approximant for bending angle $\alpha$ \label{sec:FifthOrderAlpha}}
The approximant used to compute the bending angle $\alpha$ that is used to generate the lowest-order term in the closest approach limit, $\phi_0$ (used in all figures), is as follows.  An approximant that is asymptotic to the weak field limit~(\ref{eq:alpha1}) up to 5$^{th}$ order while also being asymptotic to the strong field limit~(\ref{eq:alpha0}) is given as
\begin{equation}
 \alpha_{{\rm A}5}=-\pi+\beta+\gamma\ln\zeta+\delta_{a,1}\frac{\sqrt{3}}{b'}-\gamma\ln b'+\sum_{n=1}^{6}B_nb'^{\frac{n}{2}}\left(\Delta_{n+1}\sqrt{b'}\ln b'+\Delta_n\right)
 \label{eq:approximant5}
\end{equation}
where $\Delta_n=1+(-1)^n$,~$\delta_{a,1} = \left\{
     \begin{array}{ll}
    0 &:~ a\neq1\\
       1&:~ a=1
     \end{array}
   \right.$,

\begin{equation}\left[\begin{tabular}{l}
$B_1$ \\ 
$B_2$ \\ 
$B_3$                       \\
$B_4$                       \\
$B_5$                      \\
$B_6$
\end{tabular}     \right]=\left[
\begin{tabular}{cccccc}
$-$9/2 & 9/2       &   $-$9/2   &        9/2    &       $-$3     &       15\\      
      $-$9      &     19/2    &       $-$10   &       21/2    &       $-$6     &       45\\      
      $-$9      &       9     &       $-$9      &       9       &      0      &      60\\      
      9/2   &       $-$9/2    &        5     &       $-$6      &      12     &        0\\      
     $-$3/2     &      3/2    &      $-$3/2    &       3/2    &        3     &       15\\      
       5    &        $-$5       &      5     &      $-$9/2    &       $-$6      &      $-$45 
\end{tabular}\right]\left[\begin{tabular}{l}
$D_0$\\ 
$D_1$ \\ 
$D_2$                       \\
$D_3$                       \\
$D_4$                      \\
$D_5$
\end{tabular}     \right],
\label{eq:A5}
\end{equation}
and 
\begin{equation}
D_{j>0}=a_j+(-1)^j\left(\frac{\gamma}{j}-\delta_{a,1}\sqrt{3}\right),~D_0=\pi-\beta-\gamma\ln\zeta-\delta_{a,1}\sqrt{3}
\label{D}
\end{equation}
The constants in both~(\ref{eq:alpha0}) and the expressions above, whose functional dependence is determined entirely by the BH spin parameter $a$,  are defined as follows~\cite{Barlow:2017b}:
\begin{equation}
   \beta = \left\{
     \arraycolsep=1.4pt\def\arraystretch{2}
     \begin{array}{ll}
    0 &:~ a=0\\
       \frac{r_c^{5/2}[U_-V_-+U_+V_+]}{3\sqrt{(1-a^2)}[r_c^2-2r_c+a^2](1-a/b_c)} &:~ 0<|a|<1\\
       \frac{8\sqrt{3}-6}{9} &:~ a=-1\\
       \frac{\sqrt{3}-4}{3} &:~ a=1,
     \end{array}
   \right.
\label{eq:betaarray}
\end{equation}

\begin{equation}
   \gamma = \left\{
     \arraycolsep=1.4pt\def\arraystretch{2}
     \begin{array}{ll}
       \frac{2r_c^{3/2}\left[r_c-2\left(1-\frac{a}{b_c}\right)\right]}{\sqrt{3} [r_c^2-2r_c+a^2]\left(1-\frac{a}{b_c}\right)} &:~ -1\le a<1\\
       \frac{4}{3^{3/2}} &:~ a=1,
     \end{array}
   \right.
\label{eq:gammaarray}
\end{equation}

\begin{equation}
   \zeta = \left\{
  \arraycolsep=1.4pt\def\arraystretch{2}
     \begin{array}{ll}
       \frac{216(7-4\sqrt{3})}{\kappa} & :~-1\le a<1\\
       \frac{18}{2+\sqrt{3}} &:~ a=1.
     \end{array} 
   \right. 
\label{eq:zetaarray}
\end{equation}

\[U_\pm= \frac{3}{r_c}\left[\pm a^2 \mp 2\left(1-\frac{a}{b_c}\right)(1\pm\sqrt{1-a^2}) \pm r_c\left(1\pm\sqrt{1-a^2}-2\frac{a}{b_c}\right)\right],\]
\[V_\pm =\xi_\pm \ln\left[\frac{(1+\xi_{\pm})(1-\sqrt{3}\xi_\pm)}{(1-\xi_{\pm})(1+\sqrt{3}\xi_\pm)}\right],~ \kappa=b_c\frac{\left[3b_c\sqrt{b_c^2-a^2}-6\sqrt{3}(b_c-a)\right]}{(b_c^2-a^2)^{3/2}},\]
and
\[\xi_\pm= \sqrt{\frac{a^2}{a^2+2r_c(1\pm\sqrt{1-a^2})}},\]
where $r_c$ and $b_c$ are given by~(\ref{eq:rc}) and~(\ref{eq:bc}), respectively.

\end{document}